%% file: main.tex
\documentclass[10pt,twocolumn,letterpaper]{article}

 \usepackage[pagenumbers]{cvpr} 

\input{preamble}

\definecolor{cvprblue}{rgb}{0.21,0.49,0.74}
\usepackage[pagebackref,breaklinks,colorlinks,allcolors=cvprblue]{hyperref}

\def\paperID{937} 
\def\confName{CVPR}
\def\confYear{2025}

\title{\method: A Versatile Framework for Solving Composite and \\ Binary-Parametrised Problems on Quantum Annealers} 

\author{Natacha Kuete Meli$^1$ \hspace{.5cm} Vladislav Golyanik$^2$ \hspace{.5cm} Marcel Seelbach Benkner$^1$ \hspace{.5cm} Michael Moeller$^1$\\
$^1$University of Siegen \hspace{1cm} $^2$MPI for Informatics, SIC \\
{\tt\small www.vsa.informatik.uni-siegen.de \hspace{1cm} https://4dqv.mpi-inf.mpg.de}
}

\begin{document}

\maketitle

\input{sec/0_abstract}  
\input{sec/1_intro}
\input{sec/2_background}
\input{sec/3_method}
\input{sec/4_results}
\input{sec/5_conclusion}
{
    \small
    \bibliographystyle{ieeenat_fullname}
    \bibliography{main}
}
\input{sec/X_suppl}

\end{document}

%% file: preamble.tex
\usepackage[dvipsnames]{xcolor}

\usepackage[accsupp]{axessibility}  
\usepackage{anyfontsize}
\usepackage{graphicx}
\usepackage{tikz}
\usepackage{tikz}
\usetikzlibrary{positioning,patterns,3d,angles,quotes}
\usepackage{pgfplots}
\pgfplotsset{compat=1.18}
\usepackage{upgreek}
\usepackage{ifthen}
\usepackage{calc}
\usepackage{amsmath,amssymb,amsthm,times, bm}
\usepackage{bbm}
\usepackage{graphicx}
\usepackage{multicol} 
\usepackage{multirow}
\usepackage{caption}
\usepackage{subcaption}
\usepackage{braket}
\usepackage{pifont}
\usepackage{fontawesome}
\usepackage[ruled,linesnumbered]{algorithm2e}
\SetKwComment{Comment}{/* }{ */}

\newtheorem{lemma}{Lemma}
\usepackage{wrapfig}

\usepackage[framemethod=tikz]{mdframed}
\newcounter{algorithm}
\renewcommand{\thealgorithm}{\arabic{algorithm}}

\newenvironment{algorithmframe}[1][]{%
\refstepcounter{algorithm}%
\ifstrempty{#1}%
{\mdfsetup{frametitle={Algorithm \thealgorithm}}}%
{\mdfsetup{frametitle={Algorithm \thealgorithm: #1}}}%
\begin{mdframed}[nobreak=true]%
}{%
\end{mdframed}%
}

\newcommand{\SO}{\operatorname{SO}}

\newcommand{\N}{\mathbb{N}}
\newcommand{\R}{\mathbb{R}}
\newcommand{\B}{\mathbb{B}}
\newcommand{\Z}{\mathbb{Z}}

\newcommand{\mat}[1]{\mathbf{#1}}
\newcommand{\mcal}[1]{\mathcal{#1}}
\newcommand{\norm}[1]{|#1|}
\newcommand{\Norm}[1]{\|#1\|}
\newcommand{\method}{QuCOOP}
\DeclareMathOperator*{\argmin}{arg\,min}
\newcommand{\Matrix}[2][0.8]{%
	\setlength{\arraycolsep}{5pt}\renewcommand{\arraystretch}{#1}
	\left(\hskip-5pt\begin{array}{*{10}{c}}#2\end{array}\hskip-5pt\right)}

%% file: sec/0_abstract.tex
\begin{abstract} 
There is growing interest in solving computer vision problems such as mesh or point set alignment using Adiabatic Quantum Computing (AQC). 
Unfortunately, modern experimental AQC devices such as D-Wave only support Quadratic Unconstrained Binary Optimisation (QUBO) problems, which severely limits their applicability. 
This paper proposes a new way to overcome this limitation and introduces \method, an optimisation framework extending the scope of AQC to composite and binary-parametrised, possibly non-quadratic problems. 
The key idea of \method~is to iteratively approximate the original objective function by a sequel of local (intermediate) QUBO forms, whose binary parameters can be sampled on  AQC devices. 
We experiment with quadratic assignment problems, shape matching  and point set registration without knowing the correspondences in advance. 
Our approach achieves state-of-the-art results across multiple instances of tested problems\footnote{project page: \url{https://4dqv.mpi-inf.mpg.de/QuCOOP/}}. 
\end{abstract}

%% file: sec/1_intro.tex
\section{Introduction}
\label{sec:intro}

Adiabatic Quantum Computing (AQC) has become a promising approach for solving combinatorial optimisation problems in different fields of science including computer vision~\cite{lucas2014ising,mcgeoch2014adiabatic,shor1999polynomial,Albash_2018}. 
AQC leverages principles of quantum mechanics to explore optimisation landscapes in a way that classical methods cannot do, allowing them to tunnel through energy barriers and escape local minima. 
Unfortunately, current AQC devices such as D-Wave's machines, also referred to as \emph{Quantum Annealers}, are restricted to \emph{Quadratic Unconstrained Binary Optimisation (QUBO)} problems. 
Any deviations from the QUBO form---e.g.,~when the objective function is non-quadratic or quadratic in a non-linear function of the binary variables--- 
makes the potential benefits of AQCs not directly accessible. 
At the same time, many computer vision problems require non-quadratic objectives to obtain accurate solutions. 
Motivated by the tangible implications for computer vision of the restriction of the current AQC generation to QUBOs, we develop a new AQC-based optimisation method that can handle a certain type of non-quadratic objective functions.

\begin{figure}[t]
   \centering

   \vspace{-.2cm}   
    \includegraphics[scale=.88, trim={.4cm .1cm .4cm .2cm}, clip]{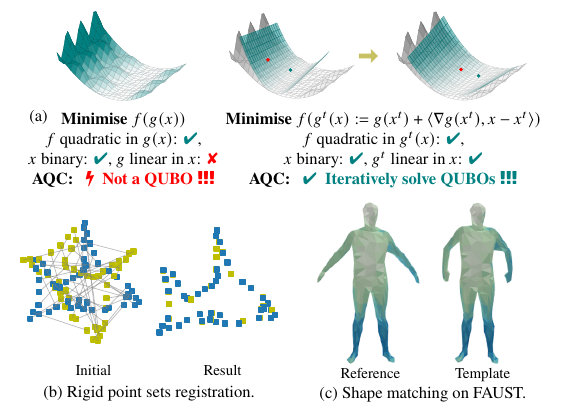}

   \vspace{-.2cm}
   \caption{(a) Our \method~framework for solving composite and binary-parametrised problems using Adiabatic Quantum Computing (AQC). 
   Some qualitative results are shown for:
   (b) Point set registration without known correspondences; 
   (c) Mesh alignment problems.
   Our general approach is competitive with specialised existing quantum and classical methods; it is compatible with quantum annealers and classical simulated annealing solvers. 
   }
   \label{fig:teaser}
\end{figure}%

We consider the general problem
\begin{equation}
\label{eq:original_problem}
    \argmin_{s \in \mcal S} \ f(s)
\end{equation}
of optimising a function $f: \mcal S \to \R$ over a feasible set \mbox{$\mcal S \subsetneq \R^n$}.
Moreover, we assume $\mcal S$ to be parameterizable by elements of a set $\mcal X \subset \R^k$ via a smooth function \mbox{$g: \mcal{X} \to \mcal{S}$} allowing to rewrite Problem~\eqref{eq:original_problem} in a composite form as
\begin{equation}
\label{eq:parametrised_problem}
    \argmin_{x \in \mcal X} \ f(g(x)).
\end{equation}
If $\mcal{X} = \R^k$, Problem \eqref{eq:parametrised_problem} describes an unconstrained problem and can be solved locally, or even globally (for $f \circ g$ convex in $x$) using gradient-based methods. 
If $\mcal{X} \subsetneq \R^k$, Problem \eqref{eq:parametrised_problem} becomes a constrained optimisation problem.
If $\mcal X$ is further non-convex, as e.g.,~the set $ \B^k := \left\{0, 1\right\}^k$ of binary vectors of length $k$, Problem \eqref{eq:parametrised_problem} is generally $\mathcal{NP}$-hard~\cite{murty1985some} and, very often, can be solved only approximately.

Gradient descent~\cite{nocedal1999numerical,nesterov2018lectures} is perhaps the most simple algorithm for solving Problem \eqref{eq:parametrised_problem} when $\mcal{X} = \R^k$.
It is an iterative method, where at each iteration $t$, a local model $f^t$ of $f$ is constructed around the current iterate $x^t$ to identify a descent direction for updating $x^t$. 
For constrained problems, one typically turns to iterative projection or additional penalty terms to ensure that all iterates $x^t$ remain feasible. 
Examples of such techniques include:
Gradient projection methods~\cite{bertsekas1976goldstein} that project the iterates of gradient descent onto $\mcal X$;
interior-point methods~\cite{potra2000interior} that minimize an augmented objective function where a so-called barrier function penalizes non-feasible iterates;
feasible direction methods~\cite{pokutta2024frank}, \eg,~Frank-Wolfe~\cite{frank1956algorithm}, that compute a feasible descent direction by solving an easier constrained problem and then update the iterate accordingly.

One of the strengths of AQC is its natural ability to optimize over discrete binary variables, whereas classical methods often resort to relaxations.
To leverage this, we present \method~(\underline{Qu}antum Framework for \underline{CO}mposite and Binary-Parametrised \underline{OP}optimisation Problems), a general framework to solve \eqref{eq:parametrised_problem} on AQCs. 
We are concerned with the case when $\mcal X = \B^k:= \left\{0, 1\right\}^k$ and $f$ is a quadratic function in $g(x)$. 
The key idea is to iteratively minimize a $1$D cut of $f$ at a current \emph{linear} Taylor approximation of $g$: 
At iteration $t$, we approximate $f$ through a local model $f^t$ by linearising the inner function $g$ around the current iterate $x^t$. 
Since $f$ is quadratic in $g(x)$ and $x$ binary, the resulting sub-objectives are QUBO problems for which modern quantum annealers provide an exciting new way to sample possible low-energy states.
As a result, our proposed optimisation does not require descent directions or step sizes in the usual sense as classical methods do, but directly targets a state $x^{t+1}$ minimising $f^t$.
See~\cref{fig:teaser}-(a) for an overview.
%

%

\paragraph{Applications.} 
Our configuration of Problem~\eqref{eq:parametrised_problem} is common in computer vision. 
Applications we are interested in this work include the Quadratic Assignment Problem (QAP,~\cite{loiola2007survey}), a reputed $\mathcal{NP}$-hard problem that aims to optimize a quadratic function over the set of permutation matrices. 
QAP models many real-life problems such as graph matching~\cite{benkner2020adiabatic,benkner2021q}, travelling salesman~\cite{gavish1978travelling}, facility location~\cite{loiola2007survey},
or scheduling problems~\cite{geoffrion1976scheduling}, 
to name just a few. 
Moreover, we show in further application examples how to solve mesh alignment \cite{benkner2021q} and rigid point set registration without input correspondences  \cite{meli2022iterative} with \method. 
See~\cref{fig:teaser}-(b, c) for representative results. 
\paragraph{Contributions.} 
Our primary technical contributions are:
\begin{itemize}
    \item \method, a new general framework for solving composite and binary-parametrised problems on AQCs (\cref{sec:method}); 
    \item An iterative scheme to construct and solve QUBO problems that locally approximate the original, perhaps non-quadratic objective function; 
    \item Applications of \method~to a variety of computer vision problems including QAP, shape alignment and point set registration (\cref{sec:qap,sec:point_set_rec}). For correspondence problems, we introduce a binary parametrisation of permutation matrices fitting in the proposed \method~framework. 
\end{itemize}
We experimentally validated the algorithm, first with simulated annealing and achieving state-of-the-art results for several considered problems (\cref{sec:results_qap,sec:results_shape_matching,sec:results_point_reg}); 
then on real D-Wave quantum annealers and improving the dimensionality of solvable permutation-related problems compared to previous quantum-compatible approaches (\cref{sec:results_dwave}).  

\section{Related Work}
\paragraph{Classical Optimisation Theory.} 
A binary optimisation technique related to ours is the quadratisation of pseudo-Boolean functions~\cite{mandal2020compressed,ishikawa2009higher}. 
Pseudo-Boolean functions are functions that can be written uniquely as a multi-linear polynomial of binary variables, with QUBOs being a particular types of those with degrees not exceeding two.
Quadratisation techniques \cite{boros2014quadratization,mandal2020compressed,rodriguez2018linear} can be generally used to reduce the degree of higher-order polynomials to two in order to obtain a QUBO form. 
However, they often require auxiliary variables, which increases the problem size and can be costly for actual AQC devices. 
In contrast, \method~does not require additional variables. 
Moreover, it is not restricted to polynomials, but could also be applied to any function as long as it fits in our composite form, see~\cref{eq:assumption}.
Next, our proposed \method~technique falls into the class of composite optimisation methods~\cite{lewis2016proximal,pauwels2016value,geiping2018composite}. 
These approaches leverage the composite form of optimisation problem~\eqref{eq:parametrised_problem} to derive easy-to-solve sub-problems that return better local optima of the original problem. 
An incomplete list of example applications of composite optimisation problems in computer vision include super-resolution from raw time-of-flight data~\cite{geiping2018composite}, image and point set registration~\cite{JanM_fair}, shape matching~\cite{burkard1998quadratic,van2011survey,benkner2021q}.  
Composite optimisation is classically well studied for continuous problems~\cite{lewis2016proximal,pauwels2016value,geiping2018composite,drusvyatskiy2019efficiency}.
This work explores similar techniques for the discrete case, especially to leverage AQC-based optimisation.
Hence, in a broad sense, this work improves our understanding of how AQCs can be used for optimising non-quadratic objectives.
\paragraph{Quantum-enhanced Computer Vision (QeCV).} 
In computer vision, AQCs have been used to solve problems such as matching involving optimising over permutation matrices~\cite{benkner2020adiabatic,benkner2021q,birdal2021quantum}; object tracking~\cite{zaech2022adiabatic} by optimising binary assignments between observations and a set of tracks; or clustering~\cite{zaech2024probabilistic} by sampling multiple binary assignments of a sample set to models its posterior distribution. 
Several other approaches~\cite{golyanik2020quantum,SeelbachBenkner2020,meli2022iterative,SeelbachBenkner2023,Bhatia2023} were focused on registering point sets. 
A feature shared by all above-mentioned methods is the construction of QUBO problems with binary solution parametrisation that suit quantum annealers.

\noindent\textbf{Remark.} QeCV methods relying on gate-based quantum computing remain substantially less explored than AQCs \cite{rathi20233d}, and many works in this domain are theoretical, heavily use simulators and are at the intersection with quantum machine learning~\cite{cong2019quantum,hur2022quantum,li2020quantum}. 
One of the reasons is that quantum annealing can be considered the most advanced quantum computing technology nowadays, while fault-tolerant quantum computers that could be useful in computer vision are under development \cite{GoogleQuantumAI2024}.

\paragraph{The IQT Approach \cite{meli2022iterative}.}
The most closely related work to ours is the iterative quantum transformation (IQT) approach of Kuete Meli \etal~\cite{meli2022iterative}. 
IQT iteratively finds rotation matrices $R$ aligning two point sets using AQC. 
Since rotations can be parametrised by non-redundant parameters $v$ (e.g., a rotation angle in $2$D and rotation angle and axis in $3$D), the original constrained problem on rotation matrices is cast in Ref.~\cite{meli2022iterative} into an unconstrained optimisation problem over rotation parameters. 
Specifically, IQT iteratively linearizes $R$ around a current rotation parameter as 
\mbox{$R(v) = R(v^t) + \braket{\nabla R(v^t), v - v^t} + \mathcal{O}(\|v - v^t\|^2)$} and uses an AQC to sample, in each iteration, a binary representation of $v$ minimising a target function quadratic in $R(v)$. 
Hence, the IQT approach can be interpreted as a special case of \method, i.e.,~a more general optimisation technique relying on AQC as will become apparent from Sec.~\ref{sec:method}. 

%% file: sec/2_background.tex
\section{Background}
\label{sec:background}
As our method is designed for AQC devices, we briefly revise their principles. 
We also review classical composite optimisation, which \method~builds upon.  
For a detailed introduction to quantum computing and numerical optimisation, see
Ref.~\cite{das2005quantum,nielsen10,sutor2019dancing} and~\cite{nocedal1999numerical,nesterov2018lectures}, respectively.  
%

\subsection{Adiabatic Quantum Computing}\label{ssec:AQC} 
\emph{Adiabatic Quantum Computing (AQC)} is a model of quantum computation which is polynomially equivalent to the circuit model \cite{aharonov2008adiabatic}. 
\emph{Quantum annealers} (QAs) are experimental realisations of the AQC paradigm that do not perfectly fulfill the adiabaticity condition
\cite{Albash_2018,transverse_field_qa_Kadowaki,das2005quantum, mcgeoch2014adiabatic}. They provide a meta-heuristic for finding globally-optimal solutions to combinatorial optimisation problems relying on the evolution of a quantum-mechanical system driven by quantum fluctuations. 
Usually they optimize QUBO objectives, i.e.,~problems of the form 
\begin{equation}
    \argmin_{x \in \B^k} \quad \braket{x, \mat Q x + \mat c},
\end{equation}
where $\braket{\cdot, \cdot}$ is the standard inner product in $\mathbb{R}^k$, $\mat Q$ a symmetric matrix, and $\mat c$ a vector. 
QUBO problems are $\mathcal{NP}$-hard in general, and solving them on a QA is of high practical interest. 
Casting computer vision problems to QUBO forms can be challenging and has become an active research area \cite{benkner2021q, birdal2021quantum, zaech2024probabilistic}. 
While QUBOs can be optimised with classical optimisation techniques such as~\emph{Simulated Annealing (SA)}~\cite{kirkpatrick1983optimization}, QA is expected to offer substantial advantages over SA in certain cases~\cite{denchev2016computational,das2005quantum}. 

In QA, the provided QUBO problem is converted to the corresponding physical Ising \emph{Hamiltonian} $\mat H_p$. 
The mapping is such that the ground state of $\mat H_p$, \ie the spin configuration of the physical system achieving the minimal energy value, encodes the solution of the QUBO. 
To obtain the ground state of $\mat H_p$, the system evolves according to a time-dependent \emph{Hamiltonian} $\mat H(t)$ changing from the initial Hamiltonian $\mat H_i$ (with a known ground state) to $\mat H_p$, e.g. in the way that 
$\mat H(t) = (1-\frac{t}{T}) \mat H_i + t \mat H_p$ for $t \in [0, T]$. 
If the system transitions according to the adiabatic theorem of quantum mechanics~\cite{Jansen_2007,kato1950,born1928beweis} it stays always in a ground state.
One of the conditions is a slow-enough transition, \eg~a one that is inversely proportional to the squared minimum (over all instantaneous $t$) \emph{spectral gap} of $\mat H$. 
In practice, modern experimental realisations of AQCs are non-adiabatic and return ground states with probabilities ${\ll}1.0$. 
To increase the probability of finding the ground state at least once, annealing is repeated multiple times.

\subsection{Composite Optimisation}
\label{sec:optimisation}
Iterative methods~\cite{nocedal1999numerical,nesterov2018lectures} solve Problem~\eqref{eq:parametrised_problem} by repeatedly minimising a local model $f^t$ built upon the actual iterate $x^t$:
\begin{equation}
    x^{t+1} := \argmin_{x \in \mathcal{X}} \quad f^t(x),
\end{equation}
where the local model $f^t$ should have some nice properties such as convexity or separability making it easier to solve.

In \emph{composite optimisation}, also known as `prox-linear' or `prox-descent'~\cite{lewis2016proximal,pauwels2016value,geiping2018composite,drusvyatskiy2019efficiency}, the idea is to exploit the underlying composite nature of the objective function in Problem~\eqref{eq:parametrised_problem} to derive a local model that, still being easier to solve, approximates the original landscape more faithfully.
Two approaches to achieve this are to either approximate the outer function $f$~\cite{geiping2018composite}, or linearize the inner function $g$
~\cite{lewis2016proximal,pauwels2016value,drusvyatskiy2019efficiency}. 
In the later case, the local model reads
\begin{align}
\label{eq:composite}
   f^t(x) := 
   f\left(g(x^t) + \Braket{\nabla g(x^t), x - x^t}\right) + \frac{1}{2\alpha^t}\Norm{x - x^t}^2_2,
\end{align}
with $\alpha^t$ being the step size that prevents $x^{t+1}$ to deviate too much from $x^t$, point around which the approximation is more reliable. 
It is shown in Ref.~\cite{pauwels2016value} that for a convex and coercive $f$ and under the Kurdyka–Łojasiewicz-property, the iterates $(x^t)_{t=1}^\infty$ for the model \cref{eq:composite} converge to a critical point, \ie a minimizer of $f$. 
We build on this later approximation, \ie the linearisation of the inner function $g$, to derive QUBOs for AQC, as we consider that the outer function $f$ is quadratic in $g(x)$. 

%% file: sec/3_method.tex
\section{The Proposed \method~Iteration}  
\label{sec:method} 

This section describes our general \method~framework, followed by its applications in \cref{sec:qap} and \cref{sec:point_set_rec}. 
Proofs of all Lemmas are given in ~\cref{sec:supp_proofs}. 

\method~is an optimisation algorithm  for composite optimisation problems of the form  \eqref{eq:parametrised_problem}, where $f$ is quadratic in $g(x)$, that iteratively uses AQC. 
The function $g: \R^k \to \R^n$ is a smooth function such that \mbox{$\mcal S = \mathrm{Im}_g(\B^k) := \left\{g(x): x \in \B^k\right\}$}, hence, $g$ parametrizes the feasible domain $\mcal S \subsetneq \R^n$ of the original Problem \eqref{eq:original_problem} by elements of the set  $\mcal X := \B^k$.
For convenience, we write Problem~\eqref{eq:original_problem} as 
\begin{equation}
\label{eq:assumption}
    f(g(x)) = \Braket{g(x), \mat Q g(x) + \mat c},
\end{equation}
for a given matrix $\mat Q \in \R^{n \times n}$ and a vector $\mat c \in \R^n$.

Note that for Problem~\eqref{eq:original_problem}, we are only interested in values of $g$ on $\B^k$.
However, from a theoretical point of view, our algorithm involves a Taylor approximation of $g$, requiring $g$ to be smooth as described above.
Such a $g$ is far from being unique. 
Moreover, in addition to $\mcal S = \mathrm{Im}_g(\B^k)$, the gradient of $g$ can take any value by design.
Thus, from a practical point of view, this function $g$ should model a ``reasonable'' or ``natural'' smooth parametrisation of $\mcal S$.

A high-level description of the algorithm is provided in~\cref{alg:our_algorithm}. 
Next, we analyse its different steps. 

\begin{algorithmframe}[\method~Iteration]
\label{alg:our_algorithm}
    Initialize $x^0$ and repeat for $t = 0, 1, 2, \ldots$
    \begin{align}
        g^t (x) &:= g(x^t) + \Braket{\nabla g(x^t), x - x^t}\label{eq:linearisation}\\
        \begin{split}
         \label{eq:minimisation_step}
        x^{t+1} &\gets \argmin_{x \in \mcal \B^k} \ \left\{f^t(x) := f(g^t(x))\right\}
            \\ &\qquad \quad \mathrm{s.t.} \quad g^t (x) \in \mcal S
        \end{split}
    \end{align}
    \vspace{-.2cm}
    Return $g(x^{t+1})$. 
\end{algorithmframe}

The first step is the familiar choice of the starting point $x^0$.
This can be a vector based on prior knowledge, the result of a previous optimisation, or any random point.
As will be discussed later in~\cref{lem:convergence}, \method~converges globally, but not necessarily to the global optimum.
Hence, different $x^0$ can lead to different results.

In a subsequent loop, at each iteration $t$, the inner function $g$ is linearised around the current iterate $x^t$ using first-order Taylor expansion,~\cref{eq:linearisation}, and the next iterate $x^{t+1}$ is computed by solving the minimisation Problem~\eqref{eq:minimisation_step}.
Evaluated on the hyperplane $g^t(x)$, $f$ results in a $k$-dimensional quadratic function $f^t$ in variable $x$.
In cases where the parametrisation $g^t(x)$ differs
substantially from the parametrisation $g(x)$, to guarantee a monotone decrease of the objective values, it might also be necessary to include a step $x^{t+1}= g^{-1}(g^t(\Tilde{x}^{t+1})) $, with $\Tilde{x}$ being the output of Problem~\eqref{eq:minimisation_step}.

\cref{fig:teaser} provides a visual intuition of the constructed local model $f^t$.
Next, we ensure the compatibility of Problem~\eqref{eq:minimisation_step} with quantum annealers (see~\cref{ssec:AQC}) through a QUBO formulation.

\paragraph{The Unconstrained Part of Problem~\eqref{eq:minimisation_step}.}
The objective function of the minimisation in Problem~\eqref{eq:minimisation_step} is already in an AQC-compatible form.
At iteration $t$, we solve the QUBO 
\begin{equation}
\label{eq:qubo}
    \argmin_{x \in \mcal X} 
    \quad \Braket{x, \mat Q^t x + \mat c^t},\;\text{with}
\end{equation}
\begin{align}
    \mat Q^t &:= \nabla g(x^t) \mat Q \nabla g(x^t)^\top,\\
    \mat c^t \ &:= \nabla g(x^t) \bigg( \mat c + 2 \mat Q \bigg( g(x^t) - \nabla g(x^t)^\top x^t\bigg) \bigg).
\end{align}
This is because the local model $f^t$ now becomes quadratic in the binary variable $x$.
Details on the derivation of this QUBO problem can be found in \cref{sec:qubo}. 
\paragraph{Meeting the Constraint of Problem~\eqref{eq:minimisation_step}.}
In classical optimisation when $\mcal X$ is convex, \cf~\cref{eq:composite}, the damping term factored to the step size ensures that $g(x)$ or $x$ remains close to $g(x^{t})$ or $x^t$. 
This is reminiscent of trust-region methods requiring that the local model $f^t$ trustfully approximates $f$, which happens in a close neighborhood of $x^t$. 
In our case, $\mcal X$ is discrete, but the quantum annealer can essentially navigate the complete $\mcal X$ domain, allowing us to explore solutions beyond the trust region. 
At the same time, as required in Problem~\eqref{eq:minimisation_step}, we have to guarantee that $g^t(x) \in \mcal S$; or, more generally, to replace it with a nearby feasible point with the objective value not much worse than the linearised estimate $f(g^t(x))$. 
As the whole \method~iteration should be carried out by AQC, ensuring this feasibility amounts to designing a penalty term in the form of a quadratic function in $g^t(x)$. 
Depending on the form of $\mcal S$, this may, or may not, be easy computationally. 
The yet resulting objective function is a standard QUBO problem that can run on AQC. 
We refer to~\cref{sec:qap,sec:point_set_rec} for exemplary penalty functions on two computer vision problems, namely graph or shape matching and point set registration. 

\paragraph{Convergence Analysis.} 
While \method~does not guarantee global convergence, it, at least, monotonically decreases the objective function values as stated next: 
\begin{lemma} 
\label{lem:convergence} 
The series $(f(g(x^t)))_{t\in \N}$ generated by~\cref{alg:our_algorithm} with reparametrisation, or without and under assumptions found in Appendix \ref{sec:supp_proofs}, monotonically decreases.
\end{lemma}

\section{Graph Matching and QAP with \method} 
\label{sec:qap}
This section shows an application of \method~to graph matching and quadratic assignment problems (QAP). 

A general graph matching problem of size $n \in \N$ can be written in matrix form as
\begin{equation}
\label{eq:qap1}
    \argmin_{\mat P \in \Pi_n} \quad \Norm{\mat A \mat P - \mat P \mat B}_F^2,
\end{equation}
where $\mat A$ and $\mat B$ are given $n \times n$ symmetric matrices and $\Pi_n$ is the set of all $n \times n$ permutation matrices~\cite{burkard1998quadratic}.
Equivalently, by assuming the constraint \mbox{$\mat P \in \Pi_n$} to be always held and ignoring resulting constant terms, ~\cref{eq:qap1} becomes
\begin{equation}
\label{eq:qap2}
    \argmin_{\mat P \in \Pi_n} \quad - \Braket{\mat A \mat P, \mat P \mat B}_F.
\end{equation}
The Quadratic Assignment Problem (QAP) aims to maximize \eqref{eq:qap2}. 
To apply \method~to \eqref{eq:qap2}, we need to parametrise the permutation matrix $\mat P$ using binary variables: 
\begin{lemma}
\label{lem:pi}
Let $((a, b), \ a, b = 1, \ldots, n,  \ a < b)$, with cardinality $k := n(n-1)/2$, be the tuple of all $2$-cycles permuting $a$ and $b$, and $\mat T_i$ the matrix representation of the $i$-th cycle. 
Any permutation matrix can be parametrised with a length-$k$ binary vector $x := (x_1, \ldots, x_k)$ via the function
\begin{align}
\label{eq:pi}
    \mat P: \B^k &\to \Pi_n \nonumber\\
    x &\mapsto \mat P(x) := \prod_{i = 1}^k \mat P_i(x_i), 
    \quad \mat P_i(x_i) := \mat T_i^{x_i},
\end{align}
where $\mat T_i^{x_i} $ is $ \mat T_i$ if $x_i = 1$ and the identity $\mat I_{n\times n}$ otherwise.
\end{lemma}

Next, we construct out of $\mat P$ from~\cref{eq:pi} a smooth function $g$ to be used in~\method.
A natural choice is to design, for each $\mat P_i$ from~\cref{eq:pi}, a smooth function defined through \mbox{$\mat P_i(x_i) = x_i (\mat T_i - \mat I_n) + \mat I_n$ for $x_i \in \R$}, which makes $\mat P_i$ and thus $\mat P$ differentiable with respect to $x_i$ and thus $x$. 
Now letting $g(x) := \mat p(x) := \text{vec}(\mat  P(x))$ be the vector formed by the rows of $\mat P(x)$, we propose to solve 
Problem~\eqref{eq:qap2} using~\cref{alg:our_algorithm} by solving at each iteration $t$ the sub-problem
\begin{equation}
\label{eq:graph_matching}
    \argmin_{x \in \mcal X} \ \braket{g^t(x), \mat Q  g^t(x)},
    \quad
    \mat Q := \alpha \mat I_{n\times n} - \mat A \otimes \mat B,
\end{equation}
where $g^t(x) = \mat p^t(x) = \mat p (x^t) + \braket{\nabla \mat p (x^t), x - x^t}$.
The term $\alpha \mat I_{n\times n}$ is, for some $\alpha \in \R_+$, the penalty added to enforce \mbox{$\mat p^t(x) \in \Pi_n$} as required by~\method.
In the next paragraph, we justify why our penalty term effectively penalizes non-feasible solutions.
Finally, the optimisation follows~\cref{alg:our_algorithm} with $x \in \mcal X = \B^k$.

\paragraph{On the Penalty Term of the Permutation Constraint.}
We next show why minimising $\alpha \Norm{\mat p^t(x)}^2_2$ effectively constrains the result of the minimisation step in \eqref{eq:graph_matching} to lie in the set $\Pi_n$ of permutation matrices.
At iteration $t$, we have 
\begin{equation}
    \mat p^t(x) = \mat p (x^t) + \braket{\nabla \mat p (x^t), x - x^t}, \label{eq:Linearisation}
\end{equation}
in which each row of $\nabla \mat p (x^t)$, that is, each partial derivative of $\mat p$ (\cf~\cref{eq:partial_p} in the supplement), contains exactly two $1$ and two $-1$ entries.
A following nice property of \method~is that its iterates $\mat P^t(x)$ partially satisfy the permutation matrix constraint in the sense that their rows and columns sum up to ones, as stated in~\cref{lem:sum_ij}.
\begin{lemma}
\label{lem:sum_ij}
All the iterates $\mat P^t(x)$ computed by~\cref{alg:our_algorithm} for $\mat P(x)$ from~\cref{eq:pi} fulfill $\mat P^t(x) \in X, \forall x \in \B^k$, where
\begin{equation}
\small
    X := \left\{ \mat M \in \Z^{n \times n} \bigg| \ \sum_{i=1}^n \mat M_{ij} = \sum_{j=1}^n \mat M_{ij} = 1, \forall ij \right\}.
\end{equation}
\end{lemma}

As a consequence of~\cref{lem:sum_ij},  \mbox{$n \leq \Norm{\mat p^t(x)}^2_2$} holds so that minimising $\Norm{\mat p^t(x)}^2_2$ enforces $\mat P^t(x) \in \Pi_n$ as desired.

\section{Point Set Registration with \method} 
\label{sec:point_set_rec}
We next apply \method~to point set registration. 
Given a reference set $\mat X:= (x_i )_{i=1}^n \in \R^{d \times n}$ and a template set $\mat Y:= ( y_i )_{i=1}^n \in \R^{d \times n}$, our goal is to determine the best rotation $\mat R \in \SO(d)$ and the best permutation $\mat P \in \Pi_n$ that jointly transform the template set and find correspondences with the reference set. 
Concretely, we aim to solve
\begin{equation}
\label{eq:rigid_problem}
    \argmin_{\substack{\mat R \in \SO(d), \ \mat P \in \Pi_n}} \quad \Norm{\mat X \mat P - \mat R \mat Y}_F^2.
\end{equation}
By assuming the constraints $\mat R \in \SO(d) $ and \mbox{$ \mat P \in \Pi_n$} to be always fulfilled and discarding resulting constant terms, we retrieve the problem formulation:
\begin{equation}
    \argmin_{\substack{\mat R \in \SO(d), \ \mat P \in \Pi_n}} 
    \quad 
    - \Braket{\mat X \mat P, \mat R \mat Y}_F.
\end{equation}

As required by~\cref{alg:our_algorithm}, we need to binary-parametrise $\mat R$ and $\mat P$.
For $\mat P$, we use the parametrisation from~\cref{lem:pi}, again with a total of $k_p := n(n-1)/2$ binary parameters. 
For $\mat R$, we follow Ref.~\cite{golyanik2020quantum,meli2022iterative} and write $\mat R$ using the exponential mapping from the orthogonal Lie algebra to its group, as a function $\mat R (y) = \exp(\mat M(y))$, where $\mat M$ is a skew-symmetric matrix depending on the rotation parameter $y \in \R^{k_r}$, with $k_r:= d(d-1)/2$.
By flattening the optimisation variables $\mat R(y)$ and $\mat P(x)$ as $\mat r(y): = \text{vec}(\mat  R(y))$ and $\mat p(x) := \text{vec}(\mat  P(x))$ our optimisation sub-problem in the \method~iteration $t$ becomes 
\begin{align}
    \argmin_{\mat (x, y) \in \mcal X} 
    \quad 
    \braket{g^t(x, y), \mat Q g^t(x, y)},\\
    \label{eq:q_psr}
        \mat Q := \Matrix{\alpha \mat I_{n \times n} & - \frac{1}{2}\mat X \otimes \mat Y \\ - \frac{1}{2}(\mat X \otimes \mat Y)^\top & \beta \mat I_{d\times d}},
\end{align}
where $g^t(x, y) := \left(\mat p^t(x), \mat r^t(y)\right)$ with the linearisation $ \mat p^t(x) = \mat p (x^t) + \braket{\nabla \mat p (x^t), x - x^t}$ for the permutation and $ \mat r^t(y) = \mat r (y^t) + \braket{\nabla \mat r (y^t), y - y^t}$ for the rotation.
The term $\alpha\mat I_{n\times n}$ ensures that the iterate $\mat P^t(x)$ is a valid permutation matrix, and $\beta\mat I_{d\times d}$ that $\mat R^t(y)$ approaches an orthonormal matrix. 
The optimisation now follows~Alg.~\ref{alg:our_algorithm} in which the permutation variable 
$x$ lies in $\B^{k_p}$, and, as in Ref.~\cite{meli2022iterative}, we binarize the rotation variable $y$ by considering its flexible $m$-bits discretisation over a tunable interval, leading to $y \in \B^{m\cdot k_r}$.
In total, $(x, y)$ is a binary vector in $\B^{k_p + m\cdot k_r}$.

%% file: sec/4_results.tex
\section{Experimental Evaluation}
\label{sec:results}

We perform numerical experiments on QAP, shape matching and point set registration problems.
\cref{sec:results_qap,sec:results_shape_matching,sec:results_point_reg} report results of our method when optimised using Simulated Annealing (SA) on classical hardware, and~\cref{sec:results_dwave} reports results of the same experiments using real D-Wave Quantum Annealer (QA).
Details about the used datasets and benchmark methods are discussed in the corresponding sections.
SA, along with the preparation of couplings and biases of \method~for the quantum annealer, is carried out on a conventional computer (AMD Ryzen 9 5900X 12-Core Processor CPU with 128GB RAM).

\paragraph{Penalty Factors.}
Similar to many other AQC methods dealing with constrained problems~\cite{lucas2014ising,benkner2020adiabatic,birdal2021quantum}, our approach is not exempt from the choice of the penalty factors $\alpha$ in \eqref{eq:graph_matching}, and $\alpha$ and $\beta$ in \eqref{eq:q_psr}. 
The penalty factors should be large enough to enforce the constraint while not significantly worsening the dynamic range of the resulting problem Hamiltonian, which otherwise could lead to incorrect solutions of the QUBO problem on the quantum annealer~\cite{mucke2025optimum}. 
In our experiments, we set the penalty factors for QAP and point set registration by grid search on smaller problems, which is a widely used approach in the field \cite{birdal2021quantum, Farina2023}. 
In the following, $\alpha$ is set for QAPs to the lowest eigenvalue for $\mat A \otimes \mat B$; for shape matching problems, $\alpha$ is ten times the lowest eigenvalue of $\mat A \otimes \mat B$; and in point set registration, $\alpha$ equals $\Norm{\mat X}_F^2$ and $\beta$ equals $0.1\cdot\Norm{\mat Y}_F^2$.

\subsection{Quadratic Assignment Problems (QAP)}
\label{sec:results_qap}
To benchmark our method on QAP, we consider the Fast Approximate Quadratic programming for graph matching (FAQ)~\cite{vogelstein2015fast} and the 2-OPT method~\cite{croes1958method} as classical competitors, and Q-Match~\cite{benkner2021q} as an AQC-compatible competitor. 
Implementations for FAQ and 2-OPT are provided within Python's Scipy-library, and our method as well as Q-Match are optimised using the SA implementation provided by D-Wave. 
Q-Match's source code is publicly available
\cite{QMatchCode2021}.
A technical comparison between \method~and Q-Match can be found in \cref{sec:supp_comparison}. 

\begin{figure}
    \vspace{-.5cm}
    \includegraphics[scale=.57, trim={0.cm 0.2cm 0cm 0cm}, clip]{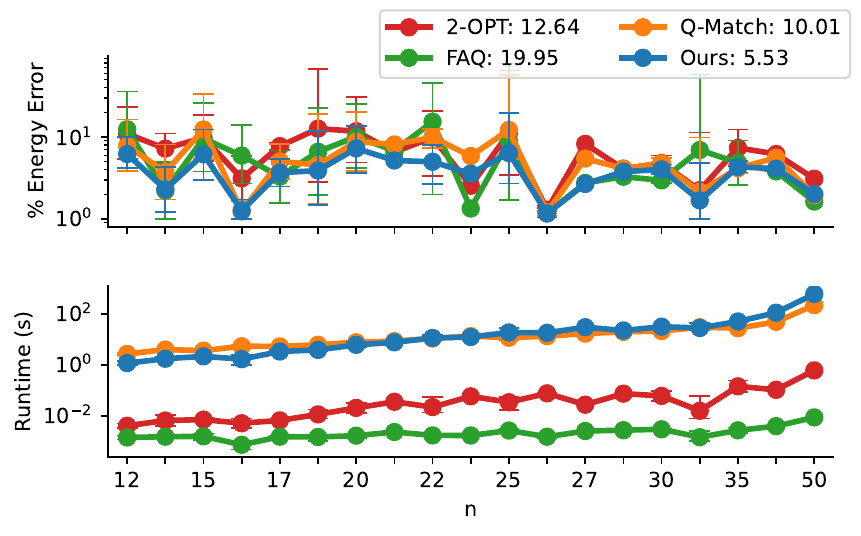}
    \vspace{-.5cm}
    \caption{Benchmark results on QAPLIB~\cite{burkard1997qaplib}.
    We plot the percentage energy error for all considered problem instances (note the means over all instances reported in the legend) and provide their runtimes. 
    Our method clearly outperforms its competitors. 
    }
    \label{fig:qaplib}
\end{figure}

We report results on the QAPLIB dataset~\cite{burkard1997qaplib} for all instances with problem size \mbox{$n\leq50$}, $72$ in total.
Details on those problem instances, along with calculated objective values, are provided in the supplement,~\cref{tab:qap}.
\cref{fig:qaplib} shows the relative errors of the computed objective values w.r.t.~optimal ones. 
Overall, our method consistently outperforms both classical and quantum competitors and has the smallest confidence intervals. 
Our method and Q-Match exhibit higher runtimes compared to FAQ and 2-Opt for almost all~$n$.
This can be explained by the fact that Q-Match solves smaller problems but requires a lot of iterations (${\approx}800$ for $n=50$), while \method~solves larger problems in much fewer iterations (${\approx}5$ for $n=50$).

\subsection{Shape Matching}
\label{sec:results_shape_matching}
We match pairs $(\mcal U, \mcal V)$ of reduced meshes from the FAUST dataset~\cite{bogo2014faust}, with $502$ vertices each following Seelbach Benkner \etal~\cite{benkner2021q}. 
We follow Ref.~\cite{benkner2021q} in several further design choices. 
Due to the high problem dimensionality, we break down each problem into carefully selected sub-problems solved iteratively. 
An initial registration based on descriptor similarities is first calculated using linear assignment.
Subsequently, in each iteration and for each vertex $u \in \mcal U$, we compute the contribution $\sum_{v \in \mcal V} \mat Q_{u\cdot n + \mat P u, v\cdot n + \mat P v}$ to the objective function, where $\mat P$ is the current solution.
The $n = 40$ vertices with the highest mismatch scores on both meshes are selected as sub-meshes and registered using either our \method~or Q-Match~\cite{benkner2021q} to enhance registration.
We repeat until the set of worst vertices converges. 
The evaluation metric is the common Princeton benchmark protocol~\cite{kim2011blended} which
calculates for each $u \in \mcal U$ the geodesic distance between the computed and the optimal vertex assignment of $u$, normalised by the diameter of $\mcal V$. 
We report cumulative geodesic errors, \ie~the percentage of assignments with error bellow incremental thresholds.

\begin{figure}[t]
\includegraphics[]{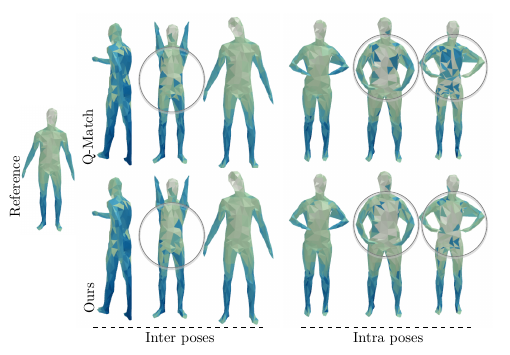}
    \vspace{-.5cm}
    \caption{Registration results on FAUST~\cite{bogo2014faust}.
    The proposed approach performs at least as well as Q-Match~\cite{benkner2021q}; see circled areas. 
    }
    \label{fig:shape_matching_experiment}
\end{figure}

\begin{figure}[t]
    \centering
    \includegraphics[scale=.55]{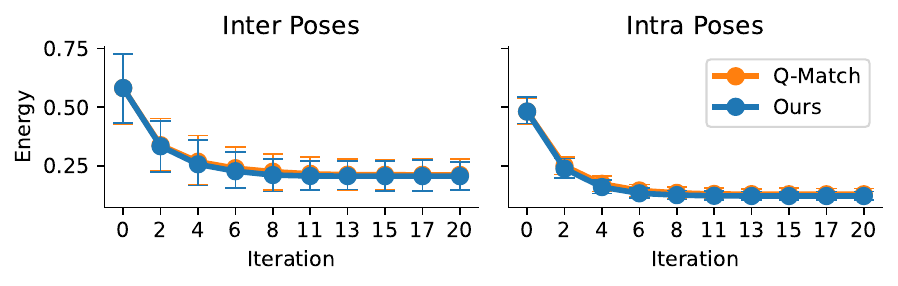}
    \includegraphics[scale=.55]{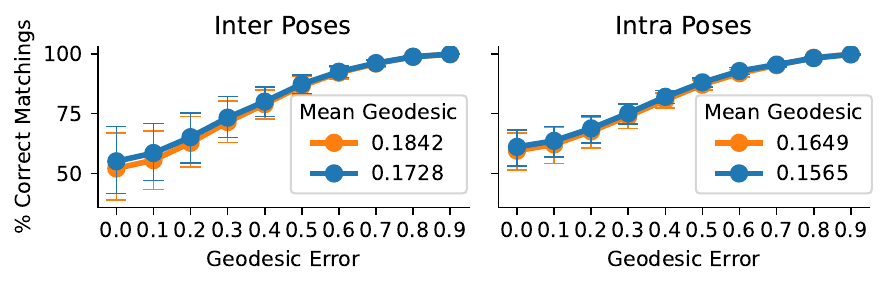}
    \vspace{-.5cm}
    \caption{Energy decrease and geodesic errors, averaged on ten instances of FAUST \cite{bogo2014faust} in each pose category.
    The mean geodesic error for both methods is given in the legend of the plot.
    The proposed method performs slightly better than Q-Match~\cite{benkner2021q}.}
    \label{fig:shape_matching_quantitative}
\end{figure}

\cref{fig:shape_matching_experiment} displays some qualitative registration results and \cref{fig:shape_matching_quantitative} presents the function value decrease over the iterations as well as the cumulative geodesic errors. 
Failure cases are discussed in~\cref{app:supp_shape_matching}. 
Our general \method~framework clearly matches the accuracy of the previous specialised Q-Match method.
It reduces the mean geodesic error by about $0.01$, as it can achieve better results on the sub-problems (see \cref{sec:supp_comparison}).
This selection of worst vertices, however, also worsens the performance of both methods, preventing them from matching the results of recent classical state-of-the-art methods~\cite{roetzer2022scalable,roetzer2024spidermatch,Zhuravlev2025}. 

\subsection{Point Set Registration}
\label{sec:results_point_reg}
We perform experiments on $2$D and $3$D point set registration.
In $2$D, points were picked on fish images, and in $3$D they were generated from geometrical curves.
The number of points varies from $n=20$ to $n=40$.
We benchmark \method~against the classical rigid CPD algorithm~\cite{Myronenko_coherent_point_drift} implemented within the pycpd package \cite{NumpyCPD2022}.
Quantum competitor methods that would have been IQT~\cite{meli2022iterative} and QA~\cite{golyanik2020quantum} cannot handle registration without correspondences.

\begin{figure}[t]
    \centering

    \begin{subfigure}{.5\textwidth}
    \hspace{-.1cm}
    \tikz
    \node[scale=.67]{
    \begin{tikzpicture}
        \node[] (A) at (0, 0) {\includegraphics[width=1.31\textwidth, trim={1.2cm 0cm .5cm 1.2cm}, clip]{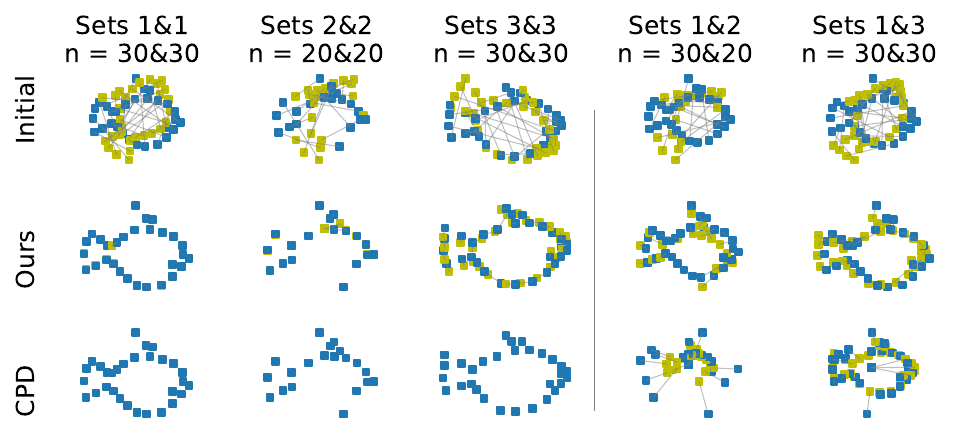}};
        \node[rotate=90, left=.0cm of A, above]{CPD \qquad Ours \qquad Initial};
		\node[fill=white, rectangle, minimum width=.5cm, minimum height=4.5cm] (B) at (1.2,0) {};
        \draw[dashed] (1.35, -2) -- (1.35, 2);
    \end{tikzpicture}
    };
    \vspace{-.2cm}
    \caption{2D}
    \end{subfigure}
    \begin{subfigure}{.5\textwidth}
    \tikz
    \node[scale=.67]{
    \begin{tikzpicture}
        \node[] (A) at (0, 0) {\includegraphics[width=1.32\textwidth, trim={1.2cm .6cm .6cm 1.2cm}, clip]{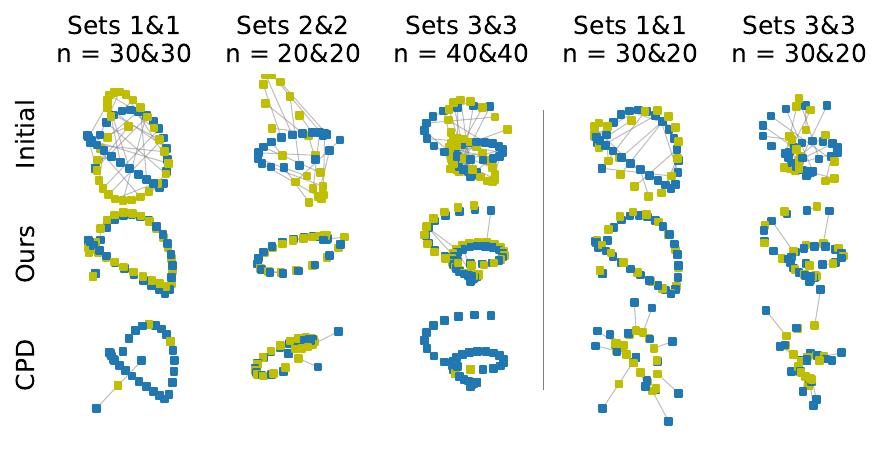}};
        \node[rotate=90, left=.0cm of A, above]{CPD \qquad Ours \qquad Initial};
		\node[fill=white, rectangle, minimum width=.5cm, minimum height=4.5cm] (B) at (1.2,0) {};
        \draw[dashed] (1.3, -2) -- (1.3, 2);
    \end{tikzpicture}
    };
    \vspace{-.2cm}
    \caption{3D}
    \end{subfigure}
    \caption{Registration results on 2D and 3D point sets with different $n$.
    Grey lines show correspondences between reference (olive) and template (blue) points.
    With up to a few correspondence errors, our method performs well on both isomorphic (left to vertical line) and non-isomorphic shapes (right to vertical line).
    CPD performs better on isomorphic shapes than on non-isomorphic shapes.}
    \label{fig:point_set_registration}
\end{figure}
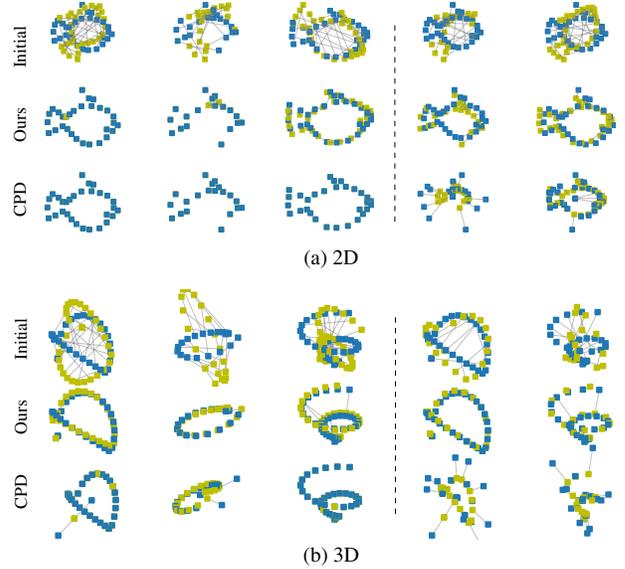

\cref{fig:point_set_registration} shows our registration results.
Left to the vertical line in the figure are point sets that are isomorphic up to the rotation and permutation, 
Right are cases where, additionally, the sets are non-isomorphic and/or have different number of points.
In the later case, we zero-pad the set with the lowest cardinality to match the other.
We use $k_r = 10$ variables for the rotation parameter and perform $15$ iterations on each problem instance. 
We see that \method's computed correspondences are overall correct, so that a reasonable transformation of the template set can still be found.
CPD shows difficulties in registering non-isomorphic sets. 

\begin{figure}
    \centering
    \includegraphics[scale=.55, trim={0.2cm .3cm 0cm 0cm}, clip]{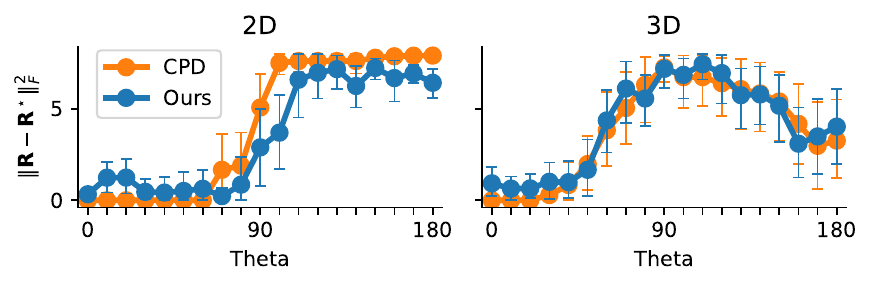}
    \vspace{-.2cm}
    \caption{Varying rotation angle in the point sets registration experiment.
    Both CPD and our method show difficulties to register the points for a rotation angle exceeding $90^\circ$ in 2D and $45^\circ$ in 3D.}
    \label{fig:point_set_registration^theta}
\end{figure}

In~\cref{fig:point_set_registration^theta}, we show registration performances, averaged over ten random problem instances of size $n=10$, for both methods, while varying the rotation angle.
The larger the rotation angle, the harder the registration for both methods. 

\subsection{Experiments on D-Wave Machines}
\label{sec:results_dwave}
\begin{figure}[t]
    \centering
    \includegraphics[scale=.65, trim={0cm .3cm .cm 0cm}, clip]{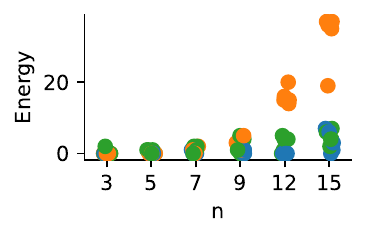}
    \includegraphics[scale=.65, trim={0cm .3cm .cm 0cm}, clip]{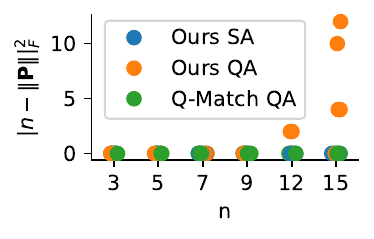}
    \vspace{-.2cm}
    \caption{D-Wave results for $5$ randomly generated QAP instances for $n = 3, 5, 7, 9, 12, 15$.
    Left are returned energy function values, $0$ being the best.
    Right is the validness $\norm{n - \Norm{\mat P}^2_F}$ of returned matrices, $0$ indicating a valid permutation matrix.
    Up to $n{\approx}12$, D-Wave finds valid permutation matrices, and often optimal ones.}
    \label{fig:dwave_qap_experiment}
\end{figure}

We now execute \method~on a real quantum device, D-Wave Advantage 6.4 accessed trough the Leap API~\cite{Dwave,Dwave_leap,Dwave_ocean}. 
Advantage has ${>}5000$ physical qubits connected in the Pegasus topology. 
However, due to the connectivity pattern of the physical qubits which induces chains in the problem embedding, it is estimated that it can solve problems of about $180$ fully connected logical variables, limiting our experiments to problems of sizes $n{\approx}15$; see also ~\cref{sec:supp_dwave}.
\paragraph{QAP.} 
On D-Wave, we solve $5$ randomly generated QAP instances for $n = 3, 5, 7, 9, 12, 15$. 
The entries of matrix $\mat A$ are $\mat A_{ij} = \Norm{x_i - x_j}_2$ for standard and normally distributed $\left\{x_i\in \R^{2}\right\}_{i=1}^n$, and $\mat B $ equals $ \mat P^\top \mat A \mat P$ such that, according to~\cref{eq:qap1}, the optimal function value is $0$. 
Each iteration of the algorithm is executed with the default annealing time of $20\mu s$ and $50$ sample reads. 
We investigate, whether the last computed matrices (\ie~$\mat P = \mat P^t(x^{t+1})$) are valid permutation matrices\footnote{Note that $\mat P^t(x^{t+1})$ may be non-valid, but $\mat P(x^{t+1})$ is always valid.}
and plot their function values.
We compare against the SA implementation and Q-Match (with QA). 
Our results are reported in~\cref{fig:dwave_qap_experiment}.
Up to $n\approx12$, \method~QA finds valid permutation matrices, but returns higher objective values compared to \method~SA and Q-Match QA. 
As $n$ becomes large, \method~QA tends to compute non-valid matrices. 
\method~QA took about $5\times$ less runtime than Q-Match QA on one problem instance, including overheads such as minor embedding and access.
Compared to previous works~\cite{benkner2020adiabatic,yurtsever2022q} which reasonably---\ie by returning valid permutation matrices---solved only problems of size $n=3, 4$ on D-Wave machines, we consider \method~as an improvement of the state-of-the-art. 

\begin{figure}
    \centering
    \includegraphics[scale=.65, trim={0.2cm .3cm 0cm 0cm}, clip]{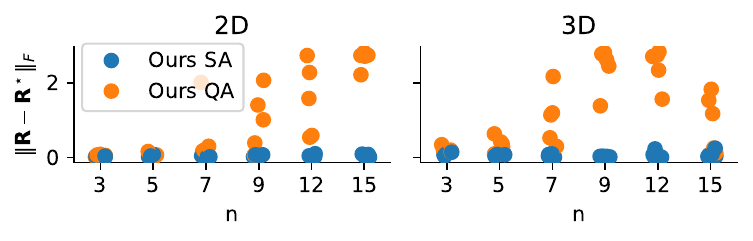}
    \vspace{-.2cm}
    \caption{D-Wave results for five randomly generated point sets registrations for $n = 3, 5, 7, 9, 12, 15$. 
    In line with the QAP experiment, D-Wave successfully registers small problem instances.}
    \label{fig:point_set_registration_n}
\end{figure}

\paragraph{Point Set Registration.} 
We perform point set registration on D-Wave for five randomly generated problem instances for $n = 3, 5, 7, 9, 12, 15$ in each set. 
The results are reported in~\cref{fig:point_set_registration_n}. 
Qualitative results can be found in \cref{sec:supp_point_set_reg}. 
We see that \method~can register the points without correspondences up to $n=5$, improving previous work~\cite{meli2022iterative} in which correspondences are known beforehand. 

%% file: sec/5_conclusion.tex
\section{Discussion and Limitations} 
\label{sec:limitations} 

The experiments reveal several features of \method: First, its ability to handle problems involving permutation problems; second, its versatility and compatibility with continuous problems such as point set registration without correspondences, which previous techniques \cite{golyanik2020quantum, benkner2021q, meli2022iterative} are not able to address. 
Moreover, \method~supports optimisation over permutation matrices up to size $n{=}12$ using the chosen parametrisation on D-Wave annealers. 

\paragraph{Approximation Accuracy.}
    As for many first-order methods, the linearisation $g^t$ of the inner function $g$ is both the key and the bottleneck of \method.
    It may not always provide accurate approximations for highly non-linear $g$ and lead to suboptimal solutions as observed in the tests. 

\paragraph{Scalability and Scaling.} 
    The space complexity of our permutation parameterisation scales quadratically with 
    $n$, 
    making it difficult to solve large problems on D-Wave, as complex embeddings increase the chain break probability.

\section{Conclusion}
\label{sec:conclusion}
We have presented \method, a framework for solving composite, binary-parametrised and possibly non-quadratic problems using AQC.
This represents a new class of problems solvable on AQC previously restricted to QUBO forms. 
\method~solves a sequence of QUBO forms approximating the original objective function, which facilitates the non-intuitive and challenging task of handcrafting QUBOs for the aforementioned problem types. 
We show that our single framework can effectively be applied to several computer vision problems. 
Namely, we address QAP and shape matching problems, outperforming the previous specialised quantum-enhanced Q-Match method by $5\%$ on QAP and $0.01$ on shape matching. 
For the first time, an AQC-compatible method for rigid point set registration is shown not to require input correspondences and match the accuracy of the classical CPD algorithm on tested problems. 
We hope this work inspires future exploration of new methods for other CV problems within the proposed framework. 
\section*{Acknowledgements}
This work was supported by the Deutsche Forschungsgemeinschaft (DFG, German Research Foundation), project number 534951134. 
%

%% file: sec/X_suppl.tex
\clearpage
\appendix
\setcounter{page}{1}

\setcounter{table}{0}
\setcounter{figure}{0}
\renewcommand{\thetable}{\Roman{table}}
\renewcommand{\thefigure}{\Roman{figure}}

\onecolumn
\begin{figure}[t]
    {\centering
    \Large
    \textbf{\thetitle}\\
    \vspace{0.5em}Supplementary Material \\
    \vspace{1.0em}}
\end{figure}

This supplementary material provides a deeper analysis of the proposed optimisation framework and more experimental details and results. 
It includes the following sections: 
\begin{itemize}
    \item Proofs of~\cref{lem:convergence,lem:pi,lem:sum_ij} claimed in \cref{sec:method} of the main text (\cref{sec:supp_proofs}); 
    \item Details of the derivation of \method's QUBO problem~\eqref{eq:qubo} of the main text (\cref{sec:qubo}); 
    \item A technical comparison between our \method~method and the Q-Match algorithm in solving permutation problems, as mentioned in  ~\cref{sec:results_qap,sec:results_shape_matching} of the main text (\cref{sec:supp_comparison}); 
    \item Details on the QABLIB results in~\cref{sec:results_qap} of the main text, as well as an investigation of the performance of an iterative local search variant of \method on QAPLIB problems (\cref{sec:supp_qaplib}); 
    \item Further experiments on shape matching from~ \cref{sec:results_shape_matching} of the main text, including failure cases on FAUST and results on the TOSCA dataset (\cref{app:supp_shape_matching}); 
    \item Qualitative results of point set registration computed on D-Wave  complementing the numerical results of~\cref{sec:results_dwave} of the main text (\cref{sec:supp_point_set_reg}); 
    \item Description of the annealing and embedding process on D-Wave's quantum annealers, relevant for the results in ~\cref{sec:results_dwave} of the main text (\cref{sec:supp_dwave}). 
\end{itemize}

\noindent
The notation in this supplement mostly follows the conventions of the main text. 
We summarise them in~\cref{notation}.

\begin{table*}[ht]
    \centering
    \begin{tabular}{p{1.5cm}p{8cm}}
    \toprule
    \textbf{Symbol} & \textbf{Meaning} \\
    \midrule
    $\mcal S \subsetneq \R^n$ & Feasible domain of an optimisation problem of interest \\
    $\mcal X = \B^k$ & Parameter set for the feasible domain $\mcal S$ \\
    $g^t$ & Linear Taylor approximation of $g$ around iterate $x^t$ \\
    $f^t$ & Local model, quadratic in $g^t$ \\
    $\nabla g$ & Gradient of $g$ with respect to $x$ \\
    $\braket{\cdot, \ \cdot}$ & Standard inner product in $\R^k$ \\
    \bottomrule
    \end{tabular}
    \caption{Main notations used in this paper and their meaning.} 
    \label{notation}
\end{table*}

\section{Proofs of the Lemmas} 
\label{sec:supp_proofs}

\paragraph{Proof of \cref{lem:convergence}.}
We show that the series of objective function values computed by \method~is non-increasing in either of the following cases: 
\begin{enumerate}
    \item \textbf{With no reparametrisation step and under following assumptions:}
    \begin{itemize}
        \item[i)] The constraint $g^t(x)\in \mathcal{S}$ is enforced with a penalty term of the form $\alpha^t  h(x, x^t) $ with sufficiently large penalty factor $\alpha^t \in \mathbb{R}_{+}$ and a function $h$ such that $h(x^t, x^t) = 0$, and $h(x, x^t) > 0$ for all $x \neq x^t$.
    
        \item[ii)] At points $g^t(x) $ in $ \mcal S$ that the algorithm can reach in a specific iteration, the linearised parametrisation $g^t$ coincides with $g$, so that $g^t(x^{t+1}) = g(x^{t+1})$.
    \end{itemize}
    
    \item 
    \textbf{Alternatively with the reparametrisation step}:
    
    It is always possible to change \cref{alg:our_algorithm} to a non-increasing algorithm by including, as mentioned in the main text, a step $x^{t+1}= g^{-1}(g^t(\Tilde{x}^{t+1})) $, where $\Tilde{x}$ is the output of Problem~\eqref{eq:minimisation_step}. 
\end{enumerate}

\begin{proof}
\begin{enumerate}

    \item \textbf{With no reparametrisation step:}

\textbf{Using Assumptions i): Majorisation-Minimisation argument:}\\
We want to show that $f(g(x^{t+1})) \leq f(g(x^t))$ for all $t$.

According to the assumption in i) the Problem~\eqref{eq:minimisation_step} can be explicitly written without constraint as
\begin{align}
    x^{t+1} &\gets \argmin_{x \in \mcal \B^k} \ f^t(x),
    \qquad
    f^t(x) := f(g^t(x)) + \alpha^t h(x, x^t)\label{eq:fgt},
\end{align}
where $\alpha^t \in \R_{>0}$ is a penalty factor and $h$ some function in $x$ ensuring the feasibility so that $g^t(x) \in \mcal S$.
At iteration $t$, the QUBO solver finds the minimiser $x^{t+1}$ of $f^t$ among all $x\in \mcal X$ with $g^t(x) \in \mcal S$.
Automatically, we have $f^t(x^{t+1}) \leq  f^t(x^t) = f(g(x^t))$.
It remains to show that
\begin{equation}
\label{eq:ultimate_goal}
    f(g(x^{t+1})) \leq f^t(x^{t+1}).
\end{equation}

We will first show  that~\cref{eq:ultimate_goal} can be established for specific designs of $h$ making $f^t$ a majoriser of $f\circ g$.
More specifically, we will establish the proof for two local models $f^t_i(x) := f(g^t(x)) + \alpha^t h_i(x), i=1,2$, with 
\begin{align}
    h_1(x, x^t) &= \braket{x - x^t, \mat H (x - x^t)}\\
    h_2(x, x^t) &= \braket{g^t(x) - g(x^t), \mat H (g^t(x) - g(x^t))},
\end{align}
where $\mat H \in \R^{k \times k}$ is some positive definite matrix ensuring the feasibility on $\mcal S$.
As a side note, the design choice $h_2$ is a special case of $h_1$ with an iteration dependent $\mat H^t$.
To see this, we plug the expression $g^t(x) = g(x^t) + \braket{\nabla g(x^t), x - x^t}$ into $h_2$, yielding $h_2(x, x^t) = \braket{\nabla g(x^t)^\top(x - x^t), \mat H \nabla g(x^t)^\top(x - x^t)}$, from which we can read out that $\mat H^t = \nabla g(x^t) \mat H \nabla g(x^t)^\top$.
Note that Assumptions i) require $ \nabla g(x^t)$ to have full rank, because otherwise $h_2(x, x^t)$ may be zero for $x \neq x^t$.

We now show that for $h$ being either $h_1$ or $h_2$, our local model $f^t$ in~\cref{eq:fgt} is a majoriser of $f \circ g$ and~\cref{eq:ultimate_goal} holds.
The proof follows the same argumentation as well-known Majorisation-Minimisation methods~\cite[Lemma 2.1 \& Theorem 2.2]{pauwels2016value}.
In essence, from the smoothness of $g$, we know from \cite[Lemma 1.2.3]{nesterov2018lectures} that there exists a Lipschitz constant $a\in \R$ such that for all $x \in \B^k$ it holds
\begin{equation}
\label{eq:g_lipschitz}
    \Norm{g(x) - g(x^t) - \braket{\nabla g(x^t), x - x^t}}_2 \leq a \Norm{x - x^t}^2.
\end{equation}
On the other hand, the quadratic function $f$ is Lipschitz continuous on the finite and bounded set $\mcal S$~\cite[Theorem 10.4]{rockafellar2015convex}, \ie~$\forall x,y\in \mcal S$ we can find $ b \in \R$ such that
\begin{equation}
\label{eq:f_quad_grow}
    \norm{f(x) - f(y)} \leq b \Norm{x - y}_2.
\end{equation}
Substituting $x = g(x)$ and $y = g^t(x)$ in~\cref{eq:f_quad_grow} and using the relation in~\cref{eq:g_lipschitz}, we get
\begin{align}
    \norm{f(g(x)) - f(g^t(x))}
    &\leq b\Norm{g(x) - g^t(x)} \\
    &\leq ab\Norm{x - x^t}^2,
\end{align}
from which follows 
\begin{equation}
\label{eq:goal}
    f(g(x)) \leq f(g^t(x)) + ab\Norm{x - x^t}^2.
\end{equation}
Next, we compare the objective in~Eqs.~\eqref{eq:fgt} and \eqref{eq:goal}.
Given that $\mat H$ is positive definite, it becomes clear that we can find a positive constant $\alpha^t$ and further majorise~\eqref{eq:goal} by $f^t_1$ as
\begin{align}
    f(g(x)) &\leq f(g^t(x)) + ab\Norm{x - x^t}^2 \\
    &\leq f(g^t(x))
    + \alpha^t \braket{(x - x^t), \mat H (x - x^t)} = f_1^t(x)\label{eq:mm1}.
\end{align}
This is also true for $h_2$.
Since $\mat H$ is positive definite and $\nabla g(x^t)$ has full rank, it follows that $\nabla g(x^t) \mat H \nabla g(x^t)$ is positive definite too. 
Hence, with a proper choice of $\alpha^t$, we can further majorise~\eqref{eq:goal} by $f_2$ as
\begin{align}
    f(g(x)) &\leq f(g^t(x)) + ab\Norm{x - x^t}^2\\
    &\leq f(g^t(x))
    + \alpha^t \braket{\nabla g(x^t)^\top(x - x^t), \mat H \nabla g(x^t)^\top(x - x^t)} = f_2^t(x)\label{eq:mm2}.
\end{align} 
Hence, by properly setting the penalty factor $\alpha^t$ we can majorise the true objective. 
Thus, $f(g(x)) \leq f^t(x)$ for all $x \in \B^k$, and specifically $f(g(x^{t+1})) + \alpha \braket{g(x^t), \mat H g(x^t)} \leq f^t(x^{t+1})$ as desired.
Looking back at the argumentation so far we notice that we never had to use any properties of $h_2 (x,x^t) $ besides the properties listed in assumption i). The assumption that $h(x^t, x^t)=0$ is crucial to obtain $f^t(x^t)= f(g(x^t))$. The second condition that $h( x, x^t ) >0 $ for all $x\neq x^t$ is important to prove that $f^t$ is a majoriser of $f \circ g$ if $\alpha_t$ is chosen big enough. The general form of the inequalities \eqref{eq:mm1} and \eqref{eq:mm2} is 
\begin{align}
    f(g(x)) &\leq f(g^t(x)) + ab\Norm{x - x^t}^2\\
    &\leq f(g^t(x))
    + \alpha^t h(x,x^t)  = f^t(x)\label{eq:mmgeneral}.
\end{align}
We can find values for $\alpha^t$ so that this holds, because 
$h(x,x^t)$ is non-zero for $x\neq x^t$ and since we only consider binary vectors for $x$ and $x^t$. For arbitrary $x$ with $x\neq x^t$ the function $h(x,x^t)$ could still have approached zero in some limit.

\textbf{Proof using Assumption ii): Local search argument:}
Since \method~optimizes over binary vectors $x$ for which $g^t(x)\in \mathcal{S} $, it follows that $g$ and $g^t$ are both valid, possibly different parametrisations of a subset of the feasible set $\mcal S$, and that we have found a better point on $\mcal S$ than $g(x^t)$.
If the parametrisation on that point differ so that $g(x^{t+1})\neq g^t(x^{t+1}) $  it is not clear why $g(x^{t+1})$ should also have a better objective value than $g(x^t)$. 
However if $g(x^{t+1}) = g^t(x^{t+1}) $ on all the relevant binary vectors $x^{t}$ so that $g^t(x^{t+1})\in  \mathcal{S}$  we have the guarantee that $g(x^{t+1})$ has also a better energy than $g(x^{t})$.

    \item 
\textbf{With the reparametrisation step:}

In the case that the two parametrisations $g^t$ and $g$ differ, as explained in the main text, we can call $\Tilde{x}$ the output of the minimisation step in Problem~\eqref{eq:minimisation_step} and re-calibrate the iterate as $x^{t+1}= g^{-1}(g^t(\Tilde{x}^{t+1}))$.
It follows that $f(g(x^{t+1})) \leq f(g^t(x^t)) = f(g(x^t))$ as desired.

\end{enumerate}
\end{proof}

Note that the re-calibration is also common in classical composite optimisation, \cf~\cite[Composite Gauss–Newton]{pauwels2016value}.
In our shape matching and point sets registration problems for the case of permutation matrices,  if the iteration step stays inside the set of permutation matrices, then our linearised parametrisation $g^t$ coincides with the original parametrisation $g$. 
We will present proof for this result later in this supplement,~\cref{lem:gt_is_perm}.
This observation saved us the recalibration step and allowed for fast convergence. In the experiments, we also noticed a monotone decrease in the objective function values.

\paragraph{Proof of \cref{lem:pi}.}
We show that any permutation matrix $\mat P$ of $n$ elements can be written as an ordered product of $k = n(n-1)/2$ binary-parametrised transpositions $\mat P_i$.

\begin{proof}    
Let $\mat P$ be an arbitrary permutation we want to decompose in this way and let $c$ be the cycle notation of $\mat P$. 
It is well known that any $c$ can be written as a decomposition of disjoint cycles: 
\begin{equation}
    c = \prod_ j f^{(j)}.
\end{equation}
We will first prove that an arbitrary cycle can be written as a decomposition in 2-cycles in a fixed order.
As the order for the 2-cycles we will w.l.o.g.~use 
\begin{equation}
    ((1,2) , (1,3),..., (1,n), (2,3) , (2,4),...., (n-1,n)   ).\label{eq:OBDA}
\end{equation}
The elements of the $\ell$-cycle we want to decompose are denoted as 
\begin{equation}
\label{eq:decomposition}
    g= (g_1, g_2 ,..., g_\ell).
\end{equation}
We construct the decomposition in an iterative fashion. 
First, let m be the index of the minimal number in the cycle so that 
\begin{equation}
    g_m = \min \{g_1,..., g_\ell \}.
\end{equation}
The following identity holds for permutations:
\begin{align}
&(g_1, g_2 ,..., g_\ell) 
= (g_m g_{m+1})( g_{m+2} , g_{m+3},..., g_\ell,g_1,..., g_{m-1 }, g_{m+1}   ). \label{eq:UsedIdentity}
\end{align}
One notices that the problem is reduced to finding the decomposition for a $(\ell-1)$-cycle with a minimal element bigger than $g_m$. 
Therefore one can repeat the procedure and the 2-cycles will be in the order given by \cref{eq:OBDA}. 

The decomposition~\eqref{eq:decomposition} only uses the elements $g_1, g_2 ,..., g_\ell$.
We can now find these decompositions for each of the disjoint cycles $f^{(j)}$ that $c$ is decomposed into. Since all the cycles $f^{(j)}$ are disjoint, the 2-cycles a cycle $f^{(i)}$ is decomposed into are disjoint from the 2-cycles a cycle $f^{(j)}$ is decomposed into for $i\neq j$. 
Since these 2-cycles are disjoint, the product commutes and can be rearranged in the order of \cref{eq:OBDA}. 
If one has another order of the cycles than the one in \cref{eq:OBDA}, one also has to apply the identity \eqref{eq:UsedIdentity} in a way that the final decomposition is in that order.

Finally, the number $k = n(n-1)/2$ of distinct cycles $c$ can be decomposed into is obvious from \cref{eq:OBDA}.
\end{proof}

We have also confirmed via computation that removing a single 2-cycle from the tuple of possible 2-cycles applied in a fixed order results in not reaching the complete $S(d)$ at least for $d\leq 6$.

\paragraph{Proof of \cref{lem:sum_ij}.}
We show that the linear approximate permutation matrices $\mat P^t(x)$ computed by our algorithm fulfil the row and column sum to $1$.

\begin{proof}
Each partial derivative of $\mat P$ can be calculated as 
\begin{align}
\label{eq:partial_p}
   \frac{\partial}{\partial x_i}   &\mat P(x) = 
    \prod_{j=1}^{i-1} \mat P_j (x_j)
    \ \ \frac{\partial}{\partial x_i} \mat P_i (x_i) 
    \prod_{j=i+1}^{k} \mat P_j (x_j)
\end{align}
with $ \frac{\partial}{\partial x_i} \mat P_i (x_i) = \mat T_i - \mat I$.
So, it is easy to see that there are exactly two rows and two columns of $\frac{\partial}{\partial x_i} \mat P(x)$ that are non-zero and contain exactly one $1$ and one $-1$ each.
Thus, for all $x \in \B^k$, it holds
\begin{align}
    \sum_{i=1}^n \braket{\nabla \mat P (x^t), x - x^t}_{ij} &= 0  \ \forall j \\
    \sum_{j=1}^n \braket{\nabla \mat P (x^t), x - x^t}_{ij} &= 0 \ \forall i.
\end{align}
As per definition $\mat P^t(x) = \mat P (x^t) + \braket{\nabla \mat P (x^t), x - x^t}$, and since $\mat P (x^t) \in \Pi_n$ fulfils the row and column sum to $1$, the validity of the claim for $\mat P^t$ is immediate.    
\end{proof}

\section{QUBO Derivation for \method}
\label{sec:qubo}
We provide details of the derivation of the QUBO in~\cref{eq:qubo}.
For ease of notation, let us first write $g^t(x)$ as
\begin{align}
    g^t (x) &= g(x^t) + \Braket{\nabla g(x^t), x - x^t} \\
    &= \underbrace{g(x^t) - \nabla g(x^t)^\top x^t}_{=:g_c} + \underbrace{\nabla g(x^t)^\top x}_{=:g_x}. 
\end{align}
Now, we have
\begin{align}
    f^t(x) &= f(g^t(x)) \\
    &= \braket{g_c + g_x , \mat Q (g_c + g_x) + \mat c} \\
    &= (g_c + g_x)^\top \mat Q (g_c + g_x) + (g_c + g_x)^\top \mat c \\
    &= g_c^\top \mat Q g_c + 2 g_x^\top \mat Q g_c + g_x^\top \mat Q g_x + g_c^\top \mat c + g_x^\top \mat c \\
    &= g_x^\top \mat Q g_x + g_x^\top (\mat c + 2 \mat Q g_c) + g_c^\top \mat Q g_c + g_c^\top \mat c,
\end{align}
where the last two terms are independent with respect to the variable~$x$.
Now taking the $\argmin$ and discarding those independent terms, we obtain 
\begin{equation}
    \argmin_{x \in \mcal X} f^t(x) = \argmin_{x \in \mcal X} g_x^\top \mat Q g_x + g_x^\top (\mat c + 2 \mat Q g_c),
\end{equation}
from which we can read out the coupling matrix $\mat Q^t$ and bias vector $\mat c$ of the QUBO in~\cref{eq:qubo}.

\section{Comparison against Q-Match}
\label{sec:supp_comparison}
An interesting question is how does \method~differs from Q-Match~\cite{benkner2021q} on permutation problems.
In Q-Match, one linearizes the inner function by selecting a set of disjoint cycles over which the optimisation is performed.
For disjoint cycles, the permutation matrix can be written as a linear function of the binary variables.
As formalised is~\cref{lem:char}, we observed that with our linearisation $\mat P^t(x) = \mat P (x^t) + \braket{\nabla \mat P (x^t), x - x^t}$ and since $\mat P (x^t) \in \Pi_n$, the minimisers $x$ are precisely those such that $x - x^t$ selects the partial derivatives of disjoint cycles in $\nabla \mat P (x^t)$, non-disjoint cycles leading to $\mat P^t(x)$ with higher Frobenius norms that are being penalised.
However, unlike Q-Match where the selection of disjoint cycles is done by hand which may be sub-optimal, our algorithm optimally selects the best set of disjoint cycles and optimises over them.
We will now give the exact characterisation which permutation matrices can be obtained with our linearisation.

\paragraph{Characterisation of Reachable Valid Permutations}

\begin{lemma}
\label{lem:char}
The matrix $\mat P^{t+1}$ obtained from $\mat P^{t}$ with~\method~ is a permutation matrix, if and only if for all 
indices where $x^{t}$ differs from $x^{t+1}$ the conjugations of the cycles $T_i^{(-1)^{x_i^{t}}}$ of the form
\begin{equation}
C_i  :=    \left(\prod_{j=1}^{i-1}  \mat T_j^{x^{t}_j} \right) \mat T_i^{(-1)^{x_i^{t}}} \left(\prod_{j=i-1}^{1}  
 \left(\mat T_j^{x^{t}_j} \right)^{-1} \right) 
\end{equation}
are disjoint.
\label{thm:disjoint}

\end{lemma}

\begin{proof}

First we observe that if we start at the identity permutation the linearisation only lands on a valid permutation if disjoint cycles are applied. Later the statement will be generalised in the above way for arbitrary $x^{t}$.
The linearisation can be written in general as 
\begin{align}
    \mat P^{t+1}(x) &= \mat P^{t}  + \sum_i (x_i- x^{t}_i)    \frac{\partial}{\partial x_i} \mat P (x)|_{x= x^{t}} 
    \left(  \mat T_i - \mat I   \right) 
    \prod_{j=i+1}^{k} \mat P_j (x^{t}_j)\\
    & =   \prod_{j=1}^{k} \mat T_j^{x^{t}_j} +  \sum_i (x_i- x^{t}_i )   \left(\prod_{j=1}^{i-1}  \mat T_j ^{x^{t}_j} \right)
    \left(  \mat T_i - \mat I   \right)
    \prod_{j=i+1}^{k} \mat T_j ^{x^{t}_j}.    \label{eq:genCasePermutation}   
\end{align}
Now we insert the zero vector for $x^{t}$. This yields 
$\mat P^{t+1}(x) =\mat I  +  \sum_i (x_i )   
     \left(  \mat T_i - \mat I   \right)$.
Furthermore, we observe that $\mat T_i - \mat I$ has $-1$ as entries on the diagonal on places where the cycle $\mat T_i$ acts non-trivially. If two cycles that are chosen are not disjoint
then one adds a $-2$ to a diagonal element of $\mat I $. Therefore, we can not obtain a permutation matrix in this case. \\

To generalise this idea we look at the general case described in \cref{eq:genCasePermutation} and divide by $\mat P^{t}$:

\begin{align}
 \left(\prod_{j=1}^{k} \mat T_j^{x_j} \right)\left(\mat P^{t}\right)^{-1}  =   \mat I    +  \sum_i (x_i- x^{t}_i )   \left(\prod_{j=1}^{i-1}  \mat T_j^{x^{t}_j} \right)
     \left(  \mat T_i - \mat I   \right)
\left(\prod_{j=i+1}^{k} \mat T_j^{x^{t}_j} \right) \left( \prod_{j=k}^{1}  \left( 
\mat T_j^{x^{t}_j} \right)^{-1}   \right) \label{eq:DivisionLastpermutation},
\end{align}
where the product symbol $\prod_{j=k}^{1}$ indicates that we want to apply the cycles in the reverse order than before. Note that the left side of the equation is only a permutation matrix if $\mat P^{t+1}$ is a permutation matrix. The right side can be further simplified:

\begin{align}
 \left(\prod_{j=1}^{k} \mat T_j^{x_j} \right)\left(\mat P^{t}\right)^{-1}  
    &=   \mat I    +  \sum_i (x_i- x^{t}_i )   \left(\prod_{j=1}^{i-1}  \mat T_j^{x^{t}_j} \right)
     \left(  \mat T_i - \mat I   \right)
      \left(\prod_{j=i}^{1}  
 \left(\mat T_j^{x^{t}_j} \right)^{-1} \right) \\ 
    &=   \mat I    +  \sum_i (x_i- x^{t}_i )   \left(\prod_{j=1}^{i-1}  \mat T_j^{x^{t}_j} \right)
     \left(  \mat T_i - \mat I   \right) \left(\mat T_i^{x_i^{t}}\right)^{-1}
     \left(\prod_{j=i-1}^{1}  
\left(\mat T_j^{x^{t}_j} \right)^{-1} \right)  \\ 
    &=   \mat I    +  \sum_i (x_i- x^{t}_i )   \left(\prod_{j=1}^{i-1}  \mat T_j^{x^{t}_j} \right)
     \left(  \mat T_i^{1-x_i^{t}} - \left(\mat T_i^{x_i^{t}}\right)^{-1}   \right) 
    \left(\prod_{j=i-1}^{1}  
 \left(\mat T_j^{x^{t}_j} \right)^{-1} \right)\\
  &=   \mat I    +  \sum_i (x_i- x^{t}_i )  (-1)^{x^{t}_i}  \left(\prod_{j=1}^{i-1}  \mat T_j^{x^{t}_j} \right)
     \left(  \mat T_i^{(-1)^{x_i^{t}}} - \mat I  \right) 
    \left(\prod_{j=i-1}^{1}  
 \left(\mat T_j^{x^{t}_j} \right)^{-1} \right) \\ 
 &=   \mat I    +  \sum_i (x_i- x^{t}_i )  (-1)^{x^{t}_i}  \left(\prod_{j=1}^{i-1}  \mat T_j^{x^{t}_j} \right)
     \left(  \mat T_i^{(-1)^{x_i^{t}}} - \mat I  \right) 
    \left(\prod_{j=i-1}^{1}  
 \left(\mat T_j^{x^{t}_j} \right)^{-1} \right)\\ 
 &=   \mat I    +  \sum_i (x_i- x^{t}_i )  (-1)^{x^{t}_i}
  \left( \left(\prod_{j=1}^{i-1}  \mat T_j^{x^{t}_j} \right) \mat T_i^{(-1)^{x_i^{t}}}  \left(\prod_{j=i-1}^{1}  
 \left(\mat T_j^{x^{t}_j} \right) ^{-1}
     \right) - \mat I  \right).
     \label{eq:last}
\end{align}

In \cref{eq:last}, we see that we have the same setting as in the special case where we started with the zero vector. There are positive binary variables $(x_i- x^{t}_i )  (-1)^{x^{t}_i}\in \{0,1\} $ that tell us if an entry changed from the last binary vector iterate. The new cycle that we consider can be obtained from the old ones through a conjugation:
\begin{equation}
C_i  :=    \left(\prod_{j=1}^{i-1}  \mat T_j^{x^{t}_j} \right) \mat T_i^{(-1)^{x_i^{t}}} \left(\prod_{j=i-1}^{1} 
 \left(\mat T_j^{x^{t}_j} \right) \right).
\end{equation}
Since conjugation does not change the cycle type, $C_i$ have the same order as the $T_i$. If $C_k,C_l$ are not disjoint and $k,l$ are indices where $x^{t+1}$ differs from $x^{t}$ then $P^{t+1} $ cannot be a permutation matrix, because in some element in the diagonal we subtract a $-2$ from the identity matrix in \cref{eq:DivisionLastpermutation}.
\end{proof}

\paragraph{Proof that the Parametrisation $g(x^t)$ Coincides with the Linearised Parametrisation $g^t(x^t)$}

\begin{lemma}
\label{lem:gt_is_perm}
Within our setting for permutation matrices
if $g^t(x^{t+1})$ is a valid permutation matrix then $g^t(x^{t+1})= g(x^{t+1})$.
\end{lemma}
\begin{proof}
The non-linearised parametrisation of the permutation matrices starting from the previous iterate $x^{t}$ is according to \cref{lem:pi}: 
\begin{align}
    g^t(x^{t+1})=\mat P^{t+1} &= \prod_{i = 1}^k    \mat T_i^{x^{t+1}_i}.
\end{align}
Using $ \mat T_i^{x_i}= \mat I + x_i (\mat T_i -\mat I )$ we obtain 
\begin{align}
    \mat P^{t+1} &= \prod_{i = 1}^k \left( 
 \mat I  + x^{t+1}_i \left(\mat T_i-\mat I \right)  \right) = \prod_{i = 1}^k \left( 
 \mat I  +  x^{t}_i \left(\mat T_i-\mat I \right) + (x^{t+1}_i-x^{t}_i ) \left(\mat T_i-\mat I \right)  \right).
\end{align}
If we factor everything out we obtain:
\begin{align}
\mat P^{t+1}
&= \prod_{i = 1}^k \left( 
 \mat I  +  x^{t}_i \left(\mat T_i-\mat I \right)\right)  + \sum_{i = 1}^k (x^{t+1}_i-x^{t}_i )  
 \prod_{j  = 1}^{i-1} \left( 
 \mat I  +  x^{t}_j \left(\mat T_j-\mat I \right)\right)\left(\mat T_i-\mat I \right) \prod_{j  = i+1}^{k} \left( 
 \mat I  +  x^{t}_j \left(\mat T_j-\mat I \right)\right) \nonumber \\&+ \textnormal{ higher order terms}.
\end{align}
This is exactly the linearised parametrisation plus some higher order terms
\begin{align}
    \mat P^{t+1} 
 &=\mat P (x^{t}) + \braket{\nabla \mat P (x^{t}), x^{t+1} - x^{t}} + \textnormal{ higher order terms}.
\end{align}
Since we are in a setting where $\mat P^{t+1}$ 
is a permutation matrix we can make use of \cref{thm:disjoint}. For 2-cycles this states that if $m,l$ are both indices where $x^{t}$ differs from $x^{t}$ then 
\begin{align}
 & \left(   \left(\prod_{j=1}^{m-1}  \mat T_j^{x^{t}_j} \right) \mat T_m^{(-1)^{x_m^{t}}} \left(\prod_{j=m-1}^{1}  
 \left(\mat T_j^{x^{t}_j} \right)^{-1} \right) - \mat I \right)
 \left(   \left(\prod_{j=1}^{l-1}  \mat T_j^{x^{t}_j} \right) \mat T_l^{(-1)^{x_l^{t}}} \left(\prod_{j=l-1}^{1}  
 \left(\mat T_j^{x^{t}_j} \right) \right) -\mat I  \right)= \mat 0.
\end{align}
The expression on the left side can be further simplified to
\begin{align}
    &\left(\prod_{j=1}^{m-1}  \mat T_j^{x^{t}_j} \right) \left(\mat T_m^{(-1)^{x_m^{t}}}  - \mat I \right)\left(\prod_{j=m-1}^{1}  
 \left(\mat T_j^{x^{t}_j} \right)^{-1} \right) 
 \left(\prod_{j=1}^{l-1}  \mat T_j^{x^{t}_j} \right)
\left(   \mat T_l^{(-1)^{x_l^{t}}} -\mat I  \right) \left(\prod_{j=l-1}^{1}  
 \left(\mat T_j^{x^{t}_j} \right)^{-1} \right) \nonumber \\
 &=\left(\prod_{j=1}^{m-1}  \mat T_j^{x^{t}_j} \right) \left(\mat T_m^{(-1)^{x_m^{t}}}  - \mat I \right)\left(\prod_{j=m}^{l-1}  \mat T_j^{x^{t}_j} \right)\left(   \mat T_l^{(-1)^{x_l^{t}}} -\mat I  \right)    \left(\prod_{j=l-1}^{1}  
 \left(\mat T_j^{x^{t}_j} \right)^{-1} \right)
 = \mat 0,
\end{align}
assuming $l>m$. Inverting all valid permutation matrices that are multiplied to the expression from the right or left side yields the equivalent equation
\begin{equation}
    \left(\mat T_m^{(-1)^{x_m^{t}}}  - \mat I \right)\left(\prod_{j=m}^{l-1}  \mat T_j^{x^{t}_j} \right)
\left(   \mat T_l^{(-1)^{x_l^{t}}} -\mat I  \right) 
 = \mat 0.
\end{equation}
Finally, 2-cycles are their own inverse and 
\begin{equation}
    \left(\mat T_m  - \mat I \right)\left(\prod_{j=m}^{l-1}  \mat T_j^{x^{t}_j} \right)
\left(   \mat T_l -\mat I  \right) 
 = \mat 0
\end{equation}
will result in terms of higher order in $x^{t+1}$ vanishing.

\end{proof}

\section{Quadratic Assignment}
\label{sec:supp_qaplib}

\paragraph{Details on QAPLIB results.}
\begin{wraptable}{r}{0.5\textwidth}
\vspace{-1cm}
\begin{center}
    \resizebox{.48\textwidth}{!}{
    \begin{tabular}{cccccccccccccccccccc}
\toprule
        $n$ & $12$ & $14$ & $15$ & $16$ & $17$ & $18$ & $20$ & $21$ & $22$  & $24$ \\
        $\#$ Instances & $8$ & $2$ & $8$ & $13$ & $2$ & $4$ & $8$ & $1$ & $3$ & $1$ \\
\midrule
        $n$ & $25$ & $26$ & $27$ & $28$ & $30$ & $32$ & $35$ & $40$ & $50$ & \multirow{2}{*}{} \\
        $\#$ Instances & $3$ & $8$ & $1$ & $1$ & $2$ & $3$ & $2$ & $1$ & $1$ & \\
\bottomrule
    \end{tabular}}
    \vspace{-.3cm}
\end{center}
    \caption{QAPLIB~\cite{burkard1997qaplib} problem sizes $n$ used in our experiments.}
    \label{tab:qaplib}
\vspace{-.5cm}
\end{wraptable}
We provide further details on the QABLIB experiment from the main text.
The number of instances per problem size is provided in \cref{tab:qaplib}.

For a better inspection of the results in \cref{fig:qaplib}, we have sorted them by problem instances, which we display in \cref{tab:qap}. 
It is clear that on most instances, our method achieves the best results among the considered benchmark methods. 
Noteworthy is that we could find optimal solutions on esc problem instances. 

\begin{table*}[t]
\begin{subtable}{1.\textwidth}
\resizebox{1.\textwidth}{!}{
{
\begin{tabular}{llllllllllllllll}
\toprule
 & nug12 & nug14 & nug15 & nug16a & nug16b & nug17 & nug18 & nug20 & nug21 & nug22 & nug24 & nug25 & nug27 & nug28 & nug30 \\
\midrule
Optimal & 578 & 1014 & 1150 & 1610 & 1240 & 1732 & 1930 & 2570 & 2438 & 3596 & 3488 & 3744 & 5234 & 5166 & 6124 \\
Ours & 612 & 1048 & 1196 & 1676 & \textbf{1262} & 1758 & 1992 & 2614 & 2554 & 3678 & 3504 & 3802 & 5368 & 5336 & \textbf{6222} \\
Q-Match & 608 & \textbf{1028} & \textbf{1182} & \textbf{1640} & 1264 & 1816 & 2014 & 2650 & 2574 & 3682 & 3608 & 3966 & 5486 & 5342 & 6408 \\
FAQ & \textbf{596} & 1054 & 1186 & 1660 & 1282 & \textbf{1742} & \textbf{1946} & \textbf{2604} & 2580 & \textbf{3632} & \textbf{3500} & \textbf{3770} & \textbf{5326} & \textbf{5284} & 6230 \\
2-OPT & 620 & 1040 & 1216 & 1704 & 1288 & 1870 & 2004 & 2738 & \textbf{2526} & 3842 & 3658 & 3866 & 5544 & 5316 & 6416 \\
\bottomrule
\end{tabular}
}}
\caption{Results on QAPLIB instances in~Ref.~\cite{nugent1968experimental}.} \label{tab:qap_1}
\end{subtable}
\begin{subtable}{1.\textwidth}
\resizebox{1.\textwidth}{!}{
{
\begin{tabular}{lllllllllllllll}
\toprule
 & chr12c & chr12b & chr12a & chr15c & chr15a & chr15b & chr18a & chr18b & chr20c & chr20b & chr20a & chr22b & chr22a & chr25a \\
\midrule
Optimal & 11156 & 9742 & 9552 & 9504 & 9896 & 7990 & 11098 & 1534 & 14142 & 2298 & 2192 & 6194 & 6156 & 3796 \\
Ours & \textbf{12978} & 11978 & 10214 & 13194 & 13486 & 10152 & 15338 & \textbf{1534} & \textbf{16288} & \textbf{2446} & \textbf{2456} & \textbf{6510} & \textbf{6376} & 4796 \\
Q-Match & 13846 & 11768 & \textbf{9916} & \textbf{12646} & \textbf{12206} & 11466 & \textbf{14466} & 1574 & 25736 & 3202 & 3026 & 6838 & 6880 & \textbf{4690} \\
FAQ & 13088 & \textbf{10468} & 33082 & 16884 & 19852 & \textbf{9112} & 15440 & 1712 & 19836 & 3206 & 3166 & 8582 & 8920 & 6744 \\
2-OPT & 14636 & 16748 & 11370 & 18634 & 14234 & 9404 & 20960 & 1710 & 28800 & 3616 & 3918 & 6810 & 7114 & 5502 \\
\bottomrule
\end{tabular}
}}
\caption{Results on QAPLIB instances in~Ref.~\cite{christofides1989exact}.} \label{tab:qap_2}
\end{subtable}
\begin{subtable}{1.\textwidth}
\resizebox{1.\textwidth}{!}{
{
\begin{tabular}{llllllllllllllll}
\toprule
 & rou12 & rou15 & esc16h & esc16i & esc16b & esc16c & esc16d & esc16j & esc16e & esc16a & esc16f & esc16g & rou20 & esc32e & esc32g \\
\midrule
Optimal & 235528 & 354210 & 996 & 14 & 292 & 160 & 16 & 8 & 28 & 68 & 0 & 26 & 725522 & 2 & 6 \\
Ours & 246244 & \textbf{368728} & \textbf{996} & \textbf{14} & \textbf{292} & \textbf{160} & \textbf{16} & \textbf{8} & \textbf{28} & \textbf{68} & \textbf{0} & \textbf{26} & \textbf{740520} & \textbf{2} & \textbf{6} \\
Q-Match & \textbf{241844} & 382094 & \textbf{996} & \textbf{14} & \textbf{292} & \textbf{160} & \textbf{16} & \textbf{8} & \textbf{28} & \textbf{68} & \textbf{0} & \textbf{26} & 762868 & \textbf{2} & \textbf{6} \\
FAQ & 245168 & 371458 & 1518 & \textbf{14} & 320 & 168 & 62 & \textbf{8} & 30 & 70 & \textbf{0} & 30 & 743884 & \textbf{2} & 10 \\
2-OPT & 242552 & 369238 & \textbf{996} & \textbf{14} & \textbf{292} & 162 & \textbf{16} & 12 & 30 & \textbf{68} & \textbf{0} & 36 & 785088 & \textbf{2} & \textbf{6} \\
\bottomrule
\end{tabular}
}}
\caption{Results on QAPLIB instances in~Ref.~\cite{eschermann1990optimized,roucairol1987sequentiel}.} \label{tab:qap_3}
\end{subtable}
\begin{subtable}{1.\textwidth}
\resizebox{1.\textwidth}{!}{
{
\begin{tabular}{llllllllllllllll}
\toprule
 & tai12a & had12 & had14 & tai15b & tai15a & had16 & tai17a & had18 & tai20a & had20 & tai25a & tai30a & tai35a & tai35b & tai40a \\
\midrule
Optimal & 224416 & 1652 & 2724 & 51765268 & 388214 & 3720 & 491812 & 5358 & 703482 & 6922 & 1167256 & 1818146 & 2422002 & 283315445 & 3139370 \\
Ours & \textbf{230704} & 1674 & 2730 & \textbf{51884360} & 402384 & 3740 & 512198 & 5432 & 730642 & 7004 & 1222504 & 1874474 & 2514120 & 296071765 & 3257058 \\
Q-Match & 233040 & \textbf{1672} & 2764 & 52057859 & 404700 & \textbf{3720} & \textbf{507218} & 5400 & 742112 & \textbf{6930} & 1222290 & 1891140 & 2567762 & \textbf{287049669} & 3291870 \\
FAQ & 244672 & 1674 & \textbf{2724} & 52028170 & \textbf{397376} & 3736 & 520696 & 5416 & 736140 & 6980 & 1219484 & \textbf{1858536} & \textbf{2460940} & 306237113 & \textbf{3227612} \\
2-OPT & 246310 & 1694 & 2742 & 51934163 & 412300 & 3750 & 523148 & \textbf{5394} & \textbf{728652} & 7016 & \textbf{1216938} & 1888344 & 2525772 & 305864564 & 3340968 \\
\bottomrule
\end{tabular}
}}
\caption{Results on QAPLIB instances in~Ref.~\cite{taillard1991robust,taillard1995comparison,hadley1992new}.} \label{tab:qap_4}
\end{subtable}
\begin{subtable}{1.\textwidth}
\resizebox{1.\textwidth}{!}{
{
\begin{tabular}{llllllllllllll}
\toprule
 & scr12 & scr15 & scr20 & bur26f & bur26a & bur26d & bur26h & bur26g & bur26e & bur26b & bur26c & kra32 & wil50 \\
\midrule
Optimal & 31410 & 51140 & 110030 & 3782044 & 5426670 & 3821225 & 7098658 & 10117172 & 5386879 & 3817852 & 5426795 & 88700 & 48816 \\
Ours & 32696 & 54926 & \textbf{111286} & 3807270 & 5446264 & \textbf{3821372} & 7131335 & 10143927 & \textbf{5388824} & \textbf{3825928} & 5430040 & \textbf{90860} & 49272 \\
Q-Match & 32758 & 54684 & 120824 & 3815606 & 5444250 & 3836955 & \textbf{7099875} & \textbf{10121633} & 5399286 & 3843293 & \textbf{5427426} & 94760 & 49900 \\
FAQ & 40758 & \textbf{53114} & 127150 & \textbf{3784562} & \textbf{5436776} & 3822209 & 7121503 & 10142604 & 5398837 & 3827015 & 5435069 & 92930 & \textbf{49126} \\
2-OPT & \textbf{31884} & 57134 & 118994 & 3793300 & 5445951 & 3823900 & 7145161 & 10121687 & 5433798 & 3844335 & 5442586 & 94360 & 49194 \\
\bottomrule
\end{tabular}
}}
\caption{Results on QAPLIB instances in~Ref.~\cite{burkard1977entwurf,scriabin1975comparison,krarup1978computer,wilhelm1987solving}.} \label{tab:qap_5}
\end{subtable}
\caption{QAP results on the QAPLIB dataset sorted by instances.
The optimal solution and the best solution among the four benchmark methods in rendered in bold.
Our method achieves the best solution most frequently on several problem instances: ``chr'', ``esc'', ``tai'' and ``bur''.
In particular, we also achieve optimal solutions on ``esc'' problem instances.
As reported in~Ref.~\cite{benkner2021q}, Q-Match performs particularly well on ``esc'' and ``had'' instances.}
\label{tab:qap}
\end{table*}

\paragraph{Iterative Local Search.}
Because our algorithm only approximates the feasible set, some feasible solutions cannot be accessed by solving the sub-problems, which results in the fact that the algorithm may not find the absolute minimiser.
We investigate the impact of iterative local search, which consists of applying some random perturbation on the actual iterate to get rid of local minima.
First, we consider multiple restarts of the algorithm with a randomly chosen starting point. 
The returned solution is the one with the lowest energy over the multiple restarts.
Second, we add some noise to the iterate $x^t$ by randomly selecting one of its entries and flipping it.
This has the effect that the objective function does not monotonically decrease any more but sometimes oscillates. 
The returned solution is that with the lowest energy over the iterations.

\Cref{tab:selected_problems} summarizes our results on $15$ selected challenging QAPLIB problem instances \cite{vogelstein2015fast}. 
The single restart version with noisy iterates performs better than the compared versions.
In practice, increasing the variance of the noise, that is, the number of iterate entries getting flipped, did not improve the results that much, and a variance set too large makes the algorithm diverge.

\begin{table}[]
\centering

\resizebox{.45\textwidth}{!}{
\begin{tabular}{lll|lll|}
\cline{4-6}
                         &                              &           & \multicolumn{3}{c|}{Number of restarts}                                                               \\ \hline
\multicolumn{1}{|l|}{\#} & \multicolumn{1}{l|}{Problem} & Optimal   & \multicolumn{1}{l|}{50 w/o noise}      & \multicolumn{1}{l|}{1 with noise}       & 1 w/o noise        \\ \hline
\multicolumn{1}{|l|}{1}  & \multicolumn{1}{l|}{chr12c}  & $11156$   & \multicolumn{1}{l|}{$\textbf{11566}$}  & \multicolumn{1}{l|}{$12470$}            & \textbf{$11566$}   \\ \hline
\multicolumn{1}{|l|}{2}  & \multicolumn{1}{l|}{rou12}   & $235528$  & \multicolumn{1}{l|}{$240664$}          & \multicolumn{1}{l|}{$\textbf{238954}$}  & $245208 $          \\ \hline
\multicolumn{1}{|l|}{3}  & \multicolumn{1}{l|}{tai15a}  & $388214$  & \multicolumn{1}{l|}{$\textbf{393476}$} & \multicolumn{1}{l|}{$400892$}           & $394642 $          \\ \hline
\multicolumn{1}{|l|}{4}  & \multicolumn{1}{l|}{chr15a}  & $9896$    & \multicolumn{1}{l|}{$\textbf{11052}$}  & \multicolumn{1}{l|}{$11562$}            & $12682  $          \\ \hline
\multicolumn{1}{|l|}{5}  & \multicolumn{1}{l|}{chr15c}  & $9504$    & \multicolumn{1}{l|}{$11758$}           & \multicolumn{1}{l|}{$12980$}            & $\textbf{11212}$   \\ \hline
\multicolumn{1}{|l|}{6}  & \multicolumn{1}{l|}{rou15}   & $354210$  & \multicolumn{1}{l|}{$367812$}          & \multicolumn{1}{l|}{$\textbf{364192}$}  & $373132$           \\ \hline
\multicolumn{1}{|l|}{7}  & \multicolumn{1}{l|}{esc16b}  & $292$     & \multicolumn{1}{l|}{$\textbf{292}$}    & \multicolumn{1}{l|}{$\textbf{292}$}     & \textbf{292}     \\ \hline
\multicolumn{1}{|l|}{8}  & \multicolumn{1}{l|}{tai17a}  & $491812$  & \multicolumn{1}{l|}{$\textbf{503140}$} & \multicolumn{1}{l|}{$508222$}           & $509066$           \\ \hline
\multicolumn{1}{|l|}{9}  & \multicolumn{1}{l|}{rou20}   & $725522$  & \multicolumn{1}{l|}{$746180$}          & \multicolumn{1}{l|}{$\textbf{734720}$}  & $751658 $          \\ \hline
\multicolumn{1}{|l|}{10} & \multicolumn{1}{l|}{chr20b}  & $2298$    & \multicolumn{1}{l|}{$2688$}            & \multicolumn{1}{l|}{$2764$}             & \textbf{2518}    \\ \hline
\multicolumn{1}{|l|}{11} & \multicolumn{1}{l|}{tai20a}  & $703482$  & \multicolumn{1}{l|}{$738124$}          & \multicolumn{1}{l|}{$\textbf{733400}$}  & $740484$          \\ \hline
\multicolumn{1}{|l|}{12} & \multicolumn{1}{l|}{chr22b}  & $6194$    & \multicolumn{1}{l|}{$6828$}            & \multicolumn{1}{l|}{$\textbf{6506}$}    & $6518$             \\ \hline
\multicolumn{1}{|l|}{13} & \multicolumn{1}{l|}{tai30a}  & $1818146$ & \multicolumn{1}{l|}{$1896112$}         & \multicolumn{1}{l|}{$1887748$}          & $\textbf{1879078}$ \\ \hline
\multicolumn{1}{|l|}{14} & \multicolumn{1}{l|}{tai35a}  & $2422002$ & \multicolumn{1}{l|}{$2537102$}         & \multicolumn{1}{l|}{$\textbf{2483626}$} & $2524962$          \\ \hline
\multicolumn{1}{|l|}{15} & \multicolumn{1}{l|}{tai40a}  & $3139370$ & \multicolumn{1}{l|}{$3277944$}         & \multicolumn{1}{l|}{$\textbf{3264446}$} & $3277034$          \\ \hline
\end{tabular}}
\caption{Randomness analysis of our algorithm on a $15$ selected QAPLIB problems.
The two digits in the problem's name stand for the problem size $n$.
Compared are three variants of the algorithm: ``$50$ w/o noise'' refers to $50$ restarts of the algorithm with different, random starting points and no noise added to the iterates; ``$1$ with noise'' refers to a single restart of the algorithm with $x^0 = \bm 0$ and noise added to the iterates; and ``$1$ w/o noise'' refers to the standard, single restart of the algorithm with $x^0 = \bm 0$ and no noise added to the iterates. 
The optimal solution and the best solution among the benchmark variants are rendered in bold.
The  ``$50$ w/o noise'' variant of the algorithm performs only slightly better than the standard run, returning five times the lowest energy compared to four for the standard version.
On the other hand, a single restart with noise added to the iterates often performs better than the two other alternatives, returning $8{\times}$ the lowest energy.}
\label{tab:selected_problems}
\end{table}

\section{Shape Matching}
\label{app:supp_shape_matching}
\paragraph{Failure Cases on FAUST \cite{bogo2014faust}.}
\begin{wrapfigure}{r}{0.5\textwidth}
    \centering
    \vspace{-1cm}
    \includegraphics[scale=1.1, trim={.2cm 0cm 0cm 0cm}, clip]{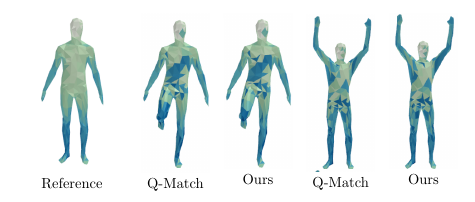}
    \vspace{-.2cm}
    \caption{
    Failure shape matching results on the FAUST dataset.
    The methods partially flip left and right, front and back vertices. 
    This particularly happens on shapes with substantial pose differences.
    }
    \vspace{-1cm}
    \label{fig:shape_matching_experiment_failure}
\end{wrapfigure}
We present failure cases of the shape matching experiment in \cref{fig:shape_matching_experiment_failure}.
Both Q-Match and our method difficulty register shapes with large pose differences. 
In particular, they cannot fix left and right, front and back vertices flip of the initial linear assignment.

\paragraph{Experiments on TOSCA Dataset.}
We perform an experiment on the TOSCA dataset \cite{bronstein2008numerical}.
We register all instances of the cat and dog classes from the dataset.

\begin{figure}[ht]
    \centering
    \begin{subfigure}{.49\textwidth}
\includegraphics[]{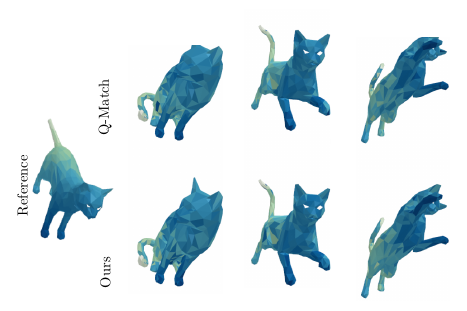}
    \vspace{-.5cm}
    \caption{Intra-classses}
    \label{fig:shape_matching_experiment_tosca_intra}
    \end{subfigure}
    \begin{subfigure}{.49\textwidth}
    \includegraphics[]{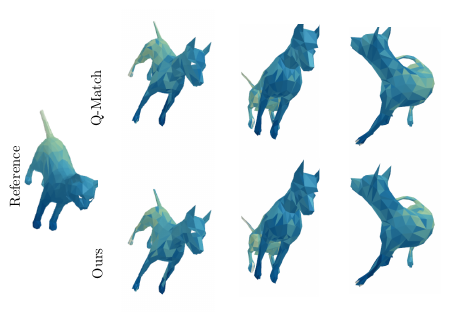}
    \vspace{-.5cm}
    \caption{Inter-classses}
    \label{fig:shape_matching_experiment_tosca_inter}
    \end{subfigure}
    \caption{Shape matching registration results on the TOSCA dataset \cite{bronstein2008numerical}. }
    \label{fig:shape_matching_experiment_tosca}
\end{figure}

In the inter-class registration, the source is a cat shape and the targets are dog shapes. 
Some qualitative results, in line with FAUST results, are presented in~\cref{fig:shape_matching_experiment_tosca_intra,fig:shape_matching_experiment_tosca_inter}.
On the TOSCA dataset also the methods are non-robust to symmetry flips.
We observed failure cases where left and right (\eg left paw registered to right one), front and back (\eg tail registered to head), were flipped, and vice versa.

\section{Point Set Registration on D-Wave}
\label{sec:supp_point_set_reg}
\begin{figure}[h]
    \centering
    \begin{subfigure}{.48\textwidth}
    \includegraphics[width=1\linewidth]{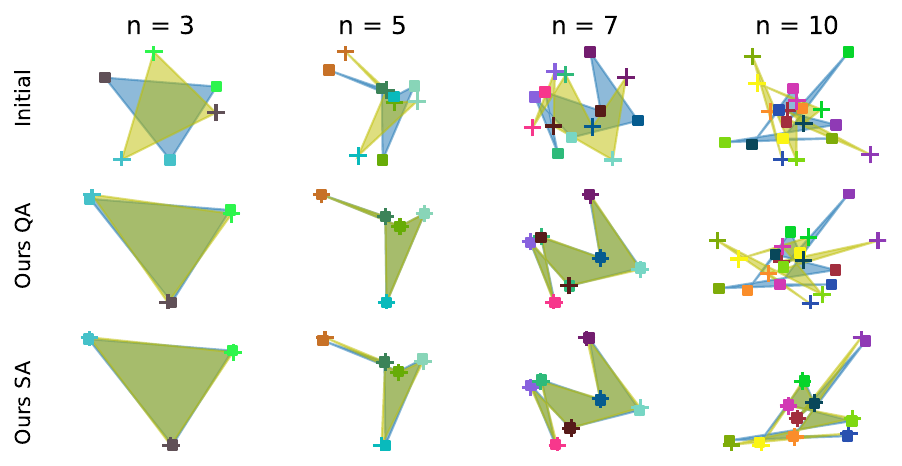}
        \caption{2D}
    \end{subfigure}
    \begin{subfigure}{.48\textwidth}
    \vspace{.5cm}
    \includegraphics[width=\linewidth]{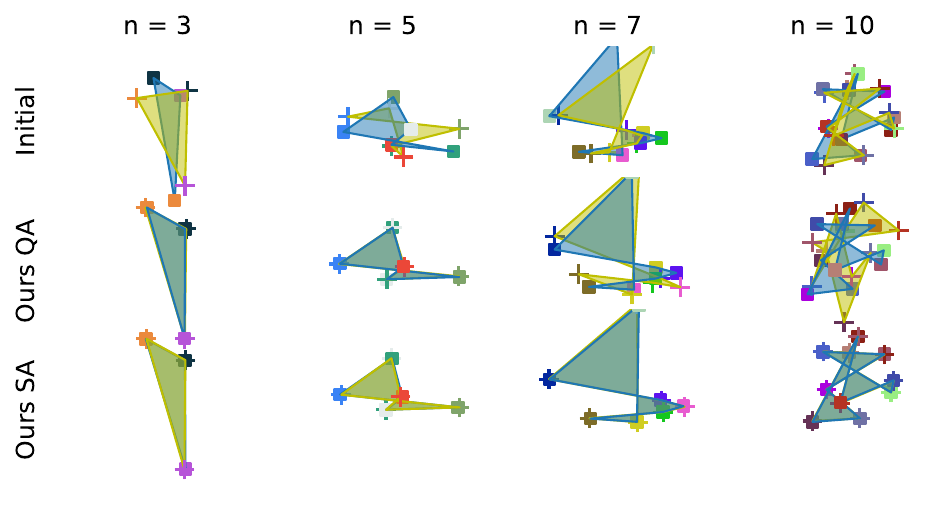}
        \caption{3D}
    \end{subfigure}
    \caption{Point set registration results computed on D-Wave.
    Shaded shapes were added for visualisation purposes only.
    Corresponding points have the same colours, while reference points are marked by squares and template points by crosses. 
    The quantum annealer can register up to $n=5$ points.
    From $n=7$, the quantum annealer can partially recover correct point assignments but fails to perfectly register the points.}
    \label{fig:dwave_point_reg}
\end{figure}

We performed point-set registration on D-Wave's quantum annealer (precisely on the Advantage system, see \cref{sec:results_dwave} for the specifics) for $n=3, 5, 7, 10$ points per set.
Our results are reported in \cref{fig:dwave_point_reg}.
Up to $n=5$ and occasionally for \mbox{$n=7$}, the quantum annealer is able to successfully register the points, and show more difficulties for larger $n$.

\begin{figure*}[ht]
    \centering
    \includegraphics[scale=.9]{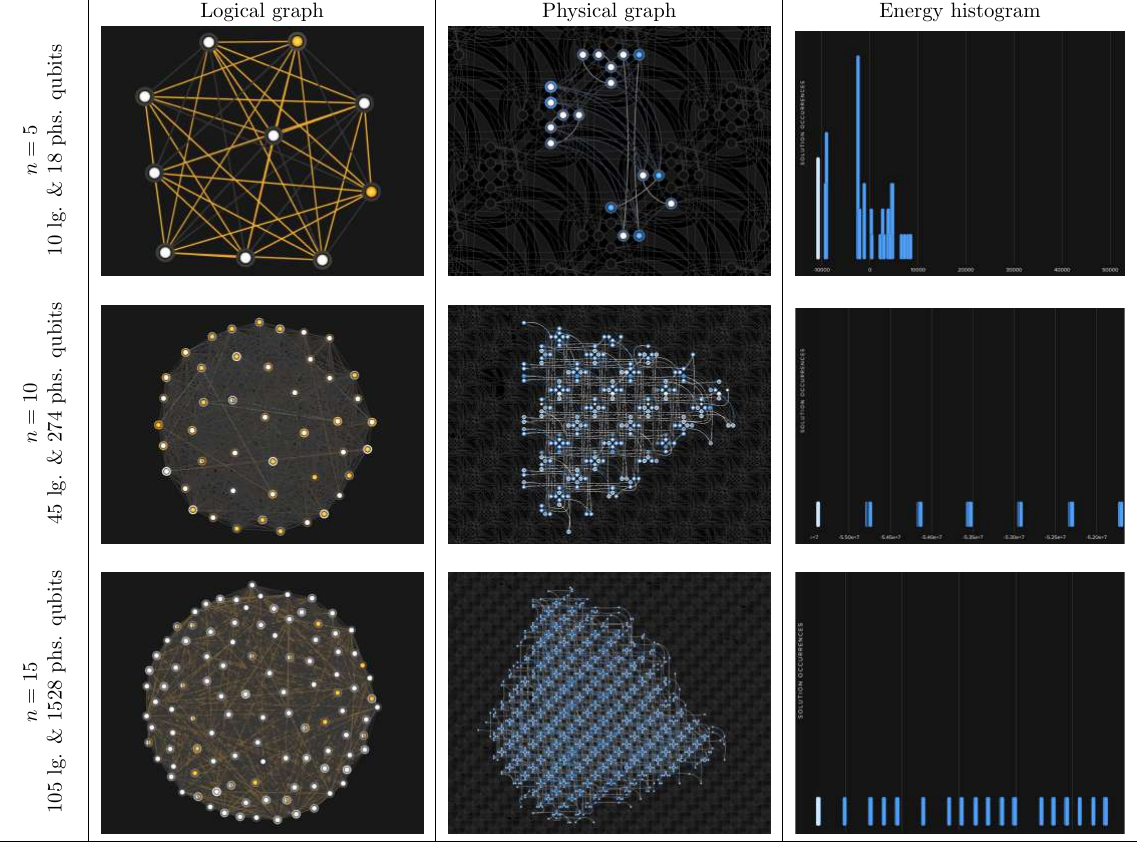}
    \caption{Minor embeddings of QAP on D-Wave for $n=5, 10$ and $15$. 
    The first column shows the logical graph of the problem, the second one visualises the physical embedding of the logical graph, and the third one is the histogram of the returned samples.} 
    \label{fig:dwave_embedding}
\end{figure*}

\section{Annealing and Embedding on D-Wave}
\label{sec:supp_dwave}
\paragraph{Annealing Process on D-Wave.}
Programming on D-Wave machines requires defining couplers and biases of the problem and sending them to a quantum processing system. 
The quantum processor creates a network of logical variables according to the problem size, which is minor-embedded in the quantum hardware. 
The network starts in a global superposition of all possible basis states. 
During the quantum annealing, the provided couplers and biases are changed into magnetic fields that deform the state landscape, emphasising the state that is most likely the solution to the underlying optimisation problem.

\paragraph{Minor Embedding.}
In most cases, our problems result in fully connected logical variables graphs.
Embedding of the logical problem onto the quantum hardware often faces the sparse variable connectivity problem. 
In order to create non-existing physical connections, the QPU \emph{chains} a set of physical variables by setting the strength
of their connecting couplers high enough to correlate them. 

\cref{fig:dwave_embedding} displays the the embedding on D-Wave of QAP for $n=5, 10, 15$, as well as the histograms of
the sampled energies.
Images are obtained from the D-Wave Leap problem inspector.
For small $n$, the histogram looks like a Boltzmann distribution concentrated around the lowest energy.
This indicates the confidence of the annealer in the solution proposal.
For larger $n$, the histogram is uniformly distributed over several different energy values. 
This is principally due to long chains, which result in chain breakages.
When this happens, the several physical variables in the chain, supposed to represent the same logical variable, get discordant spin configurations, which disturb the overall energy of the system.
\begin{wrapfigure}{r}{0.5\textwidth}
\centering
    \includegraphics[width=0.8\linewidth]{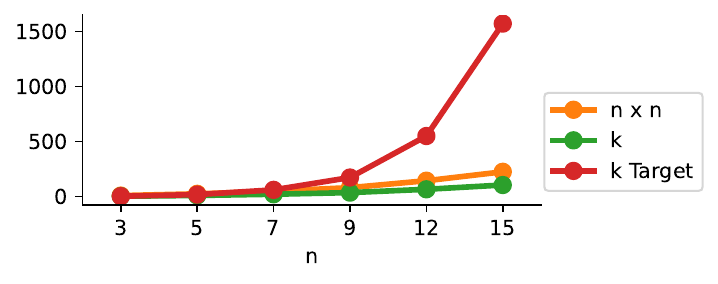}
    \caption{Comparing the problem dimension $n \times n$, the number of logical variable $k = n(n-1)/2$ of our problem modelling and the number of corresponding target variables ``$k \ \text{Target}$'' allocated on the D-Wave machine for different $n$ on QAP.}
    \label{fig:dwave^target_num_var_experiment}
\end{wrapfigure}
We tried setting the chain strength higher to avoid chain breakages. 
While this successfully eliminates chain breakages, we observed that it worsened the overall result of the algorithm.

\cref{fig:dwave^target_num_var_experiment} presents, for QAP problem instances, the growth of the number of logical and physically allocated variables as a function of the problem size $n$. 
The number of allocated physical variables is about ten times the number of logical variables.